\documentclass[12pt,letterpaper]{article}
\usepackage{amsmath,amssymb,array,calc,rotating,epsfig,psfrag,amscd, cite}

\makeatletter
\renewcommand\section{\@startsection {section}{1}{\z@}%
                                   {-3.5ex \@plus -1ex \@minus -.2ex}
                                   {2.3ex \@plus.2ex}%
                                   {\normalfont\large\bfseries}}
\renewcommand\subsection{\@startsection{subsection}{2}{\z@}%
                                     {-3.25ex\@plus -1ex \@minus -.2ex}%
                                     {1.5ex \@plus .2ex}%
                                     {\normalfont\bfseries}}
\makeatother

\let\non\nonumber

\let\l=\lambda
\let\r=\rho
\let\s=\sigma

\newcommand{\bea}{\begin{eqnarray}}
\newcommand{\eea}{\end{eqnarray}}
\newcommand{\be}{\begin{equation}}
\newcommand{\ee}{\end{equation}}

\newcommand{\p}{\partial}

\newcommand{\C}[1]{$(\ref{#1})$}

\def\IZ{\relax\ifmmode\mathchoice
{\hbox{\cmss Z\kern-.4em Z}}{\hbox{\cmss Z\kern-.4em Z}}
{\lower.9pt\hbox{\cmsss Z\kern-.4em Z}} {\lower1.2pt\hbox{\cmsss
Z\kern-.4em Z}}\else{\cmss Z\kern-.4em Z}\fi}
\def\IR{\relax{\rm I\kern-.18em R}}

\def\one{{\hbox{ 1\kern-.8mm l}}}

\newlength{\bredde}
\def\slash#1{\settowidth{\bredde}{$#1$}\ifmmode\,\raisebox{.15ex}{/}
\hspace*{-\bredde} #1\else$\,\raisebox{.15ex}{/}\hspace*{-\bredde}
#1$\fi}

\newsavebox{\zzzbar}
\sbox{\zzzbar}
  {\setlength{\unitlength}{0.9em}
  \begin{picture}(0.6,0.7)
  \thinlines
  \put(0,0){\line(1,0){0.6}}
  \put(0,0.75){\line(1,0){0.575}}
  \multiput(0,0)(0.0125,0.025){30}{\rule{0.3pt}{0.3pt}}
  \multiput(0.2,0)(0.0125,0.025){30}{\rule{0.3pt}{0.3pt}}
  \put(0,0.75){\line(0,-1){0.15}}
  \put(0.015,0.75){\line(0,-1){0.1}}
  \put(0.03,0.75){\line(0,-1){0.075}}
  \put(0.045,0.75){\line(0,-1){0.05}}
  \put(0.05,0.75){\line(0,-1){0.025}}
  \put(0.6,0){\line(0,1){0.15}}
  \put(0.585,0){\line(0,1){0.1}}
  \put(0.57,0){\line(0,1){0.075}}
  \put(0.555,0){\line(0,1){0.05}}
  \put(0.55,0){\line(0,1){0.025}}
  \end{picture}}

\newcommand{\ena}{\end{eqnarray}}
\newcommand{\beqa}{\begin{eqnarray}}
\newcommand{\eeqa}{\end{eqnarray}}

\def\l{\lambda}

\def\r{\rho}
\def\s{\sigma}

\begin{document}
\begin{titlepage}

\begin{center}



\vskip 2 cm
{\Large \bf Constraining non--BPS interactions from counterterms in three loop maximal supergravity}\\
\vskip 1.25 cm { Anirban Basu\footnote{email address:
    anirbanbasu@hri.res.in} } \\
{\vskip 0.5cm Harish--Chandra Research Institute, Chhatnag Road, Jhusi,\\
Allahabad 211019, India\\}

\end{center}

\vskip 2 cm

\begin{abstract}
\baselineskip=18pt

The structure of one, two and three loop counterterms imposes strong constraints on several non--BPS interactions in the low momentum expansion of the three loop four graviton amplitude in maximal supergravity. The constraints are imposed by demanding consistency with string amplitudes. We analyze these constraints imposed on the $D^8\mathcal{R}^4$ interaction in 11 dimensional supergravity compactified on $T^2$. We also discuss partial contributions to interactions at higher orders in the momentum expansion.

\end{abstract}

\end{titlepage}


\section{Introduction}

Obtaining the effective action of string theory in various backgrounds is useful from the low energy perspective. It is also useful in order to get a detailed quantitative understanding of the perturbative and non--perturbative duality symmetries of string theory. The effective action which encodes S matrix elements of the theory is manifestly duality invariant\footnote{For self--dual theories, what we want are duality covariant equations of motion in the given background. }. It contains both local and non--local terms, the later coming from integrating out massless modes in the loop diagrams. While calculating the effective action in arbitrary backgrounds is rather complicated, several terms have been obtained for the case of toroidal compactifications preserving maximal supersymmetry~\cite{Green:1997tv,Green:1997as,Kiritsis:1997em,Green:1998by,Obers:1998fb,Green:1999pu,Green:2005ba,Basu:2007ru,Basu:2007ck,Basu:2008cf,Green:2010kv,Green:2010wi,Basu:2011he,Gubay:2014lwa,Bossard:2014lra}. While a class of BPS interactions have been understood in detail, non--BPS interactions have been hardly analyzed.  

Maximal supergravity has played an important role in determining these interactions because of the ability to calculate multiloop amplitudes~\cite{Green:1982sw,Green:1997as,Bern:1998ug,Green:1999pu,Bern:2007hh,Green:2008bf,Bern:2008pv,Bern:2009kd,Bern:2010ue,Basu:2014hsa}. In particular, the one and two loop amplitudes have yielded several interactions in the effective action. At three loops the leading interaction is the $1/8$ BPS $D^6\mathcal{R}^4$ interaction, whose moduli dependent coefficient is highly constrained. The non--BPS $D^8\mathcal{R}^4$ interaction is the first subleading interaction in the low momentum expansion of the three loop four graviton amplitude. We perform a simple analysis to determine the constraints counterterms impose on the moduli dependent coefficient of this interaction. To do so, we isolate the one, two and three loop ultraviolet divergences of this three loop amplitude which have to be cancelled in the quantum theory by local counterterms. The structure of these counterterms is highly constrained by the structure of string theory, which uniquely fixes their renormalized values. Thus demanding the cancellation of the divergences gives us several finite contributions to the $D^8\mathcal{R}^4$ interaction. This includes contributions that are inconsistent with the structure of perturbative string amplitudes at various genera. Hence these contributions must cancel completely in the final answer, which includes the finite supergravity contributions from three loops, as well as finite and regularized contributions from higher loops. Thus we obtain an intricate interplay between the cancellation of divergences and string perturbation theory, which sheds light on the structure of quantum supergravity. We also consider some simple counterterm contributions to the $D^{10}\mathcal{R}^4, D^{12}\mathcal{R}^4, D^{14} \mathcal{R}^4$ and $D^{16}\mathcal{R}^4$ interactions which we regularize. Our analysis leads to perturbative contribution to string amplitudes at various genera. We perform the calculations in 11 dimensional supergravity on $T^2$ to be concrete, though our analysis can be generalized to arbitrary toroidal compactifications.  

Our analysis clearly shows the complications involved in the analysis of non--BPS interactions compared to their BPS counterparts. Not only are they expected to receive contributions from all loops in supergravity, but also their contributions at at every loop order are more involved than the BPS ones. Such interactions have not been studied in detail\footnote{See~\cite{Green:2006gt,Green:2008bf,Bern:2009kd,Bjornsson:2010wm,Bjornsson:2010wu,Basu:2013goa,Basu:2013oka} for some analysis of non--BPS interactions.}, and are crucial in understanding the effective action beyond the first few orders in the low momentum expansion.    

We begin by reviewing the various relations expressing quantities in M theory compactified on $T^2$ in terms of the moduli of the type IIB theory~\cite{Hull:1994ys,Witten:1995ex,Aspinwall:1995fw,Schwarz:1995jq}. 
Keeping only the scalars and the graviton obtained from the $T^2$ compactification of M theory where $R_{11}$ and $R_{10}$ are the dimensionless radii (in units of $l_{11}$) of the two circles and dropping the 1 form gauge potentials for simplicity, the line element in M theory is given by
\be \label{metric} ds^2 = G_{MN} dx^M dx^N = G^{(9)}_{\mu\nu} dx^\mu dx^\nu + R_{11}^2 l_{11}^2(dx^{11} - C dx^{10})^2 + R_{10}^2 l_{11}^2dx_{10}^2,\ee
where $x^{11}$ and $x^{10}$ are dimensionless angular coordinates.
The 9 dimensional metric $G^{(9)}_{\mu\nu} = g_{\mu\nu}^{B}$ where $g_{\mu\nu}^B$ is the type IIB metric in the string frame. The dimensionless volume $\mathcal{V}_2$ (in units of $4\pi^2 l_{11}^2$) and the complex structure $\Omega$ of the $T^2$ are related to the type IIB moduli by the relations
\be \label{rel1}\mathcal{V}_2 = R_{10} R_{11} = e^{\phi^B/3} r_B^{-4/3} , \quad \Omega_1 = C_0 , \quad \Omega_2 = \frac{R_{10}}{R_{11}} = e^{-\phi^B} , \ee 
where  $\phi^B$ is the type IIB dilaton, and $r_B$ is the dimensionless radius of the tenth dimension (in units of $l_s$) in the type IIB string frame metric given by
\be \label{rel2} r_B = \frac{1}{R_{10} \sqrt{R_{11}}},\ee
and $C_0$ is the type IIB 0 form potential. This enables us to express the interactions in M theory on $T^2$ in terms of type IIB interactions on $S^1$. 

The T dual type IIA theory has metric and moduli given by
\be g_{\mu\nu}^{B} = g_{\mu\nu}^{A}, \quad r_B = r_A^{-1} , \quad e^{-\phi^B} = r_A e^{-\phi^A}, \quad C_0 = C_1,\ee
where $g_{\mu\nu}^A$ is the type IIA metric in the string frame, $r_A$ is the dimensionless radius of the tenth dimension (in units of $l_s$) in this metric, and $C_1$ is the 1 form potential. Thus the results we obtain for the type IIB theory can be easily converted to results for the type IIA theory. 

Finally, the 11 dimensional Planck length is related to the string length by the relation
\be  l_{11} = e^{\phi^B/3} r_B^{-1/3}l_s. \ee 

In the next section, we discuss one and two loop counterterms needed to cancel one and two loop ultraviolet divergences in the four graviton amplitude. We then discuss the general structure of three loop divergences which is the main focus of the paper. The various counterterm contributions to the $D^8\mathcal{R}^4$ interaction are discussed in detail in the following section, based on the structure of the three loop four graviton amplitude. Contributions from loop diagrams involving both the ladder and Mercedes skeletons are considered, which lead to several finite terms in the effective action. Consistency with string perturbation theory imposes severe constraints on which terms can survive in the effective action. As a result certain terms which seem to survive in the effective action based on our analysis must vanish in the amplitude when all other contributions are taken into account. Thus string theory plays a decisive role in regulating the ultraviolet divergences in a way consistent with string duality. Details of some of the calculations are mentioned in the appendices.

\section{Counterterms for one and two loop ultraviolet divergences}

At one loop, the leading correction to the Einstein--Hilbert action is the $1/2$ BPS $\mathcal{R}^4$ interaction. The $\mathcal{R}^4$ interaction is $\Lambda^3$ divergent~\cite{Green:1982sw,Fradkin:1982kf,Green:1997as,Russo:1997mk,Green:1997ud}. This leads to a term in the effective action of the form
\be l_{11}^{-1} \int d^9 x \sqrt{-G^{(9)}} \mathcal{V}_2  \mathcal{R}^4 (\Lambda l_{11})^3,\ee
leading to the interaction 
\be l_s^{-1} \int d^9 x \sqrt{-g^{B}} r_B^{-1} \mathcal{R}^4 (\Lambda l_{11})^3\ee
in the type IIB effective action which receives perturbative contribution at genus one.
This is cancelled by a one loop counterterm with coefficient $c_1$ given by
\be \label{defc}\frac{4\pi^{5/2}}{3} (\Lambda l_{11})^3 + c_1 = 4\zeta(2)\ee
leaving behind a finite remainder in the M theory effective action. 
The next interaction in the low energy expansion is the $D^4 \mathcal{R}^4$ interaction which is finite. Hence \C{defc} is the only one loop counterterm (along with counterterms for interactions in the same supermultiplet) which leaves a finite remainder for terms in the effective action.   

At two loops~\cite{Bern:1998ug,Green:1999pu,Green:2008bf}, the leading interaction is the $1/4$ BPS $D^4 \mathcal{R}^4$ interaction which has a $\Lambda^8$ primitive divergence. The other divergent interactions are the $D^6 \mathcal{R}^4, \ldots, D^{12} \mathcal{R}^4$ interactions which have $\Lambda^6, \ldots, {\rm ln}\Lambda$ primitive divergences respectively. Thus the primitive divergences lead to terms in the effective action of the form
\be l_{11}^{2n-1} \int d^9 x \sqrt{-G^{(9)}} \mathcal{V}_2 D^{2n} \mathcal{R}^4 (\Lambda l_{11})^{12-2n}\ee
for $2 \leq n \leq 5$, and
\be  l_{11}^{11} \int d^9 x \sqrt{-G^{(9)}} \mathcal{V}_2 D^{12} \mathcal{R}^4 {\rm ln}(\Lambda l_{11}). \ee
Consequently, these lead to terms in the effective action of the type IIB theory given by
\bea l_s^{2n-1} \int d^9 x \sqrt{-g^{B}} r_B^{-2n/3-1} e^{2n\phi^B/3} D^{2n} \mathcal{R}^4 (\Lambda l_{11})^{12-2n},\eea
for $2 \leq n \leq 5$, and
\be l_s^{11} \int d^9 x \sqrt{-g^{B}} r_B^{-5} e^{4\phi^B} D^{12} \mathcal{R}^4 {\rm ln}(\Lambda l_{11}).\ee

The primitive divergences for the $D^4\mathcal{R}^4$, $D^8\mathcal{R}^4$ nd $D^{10} \mathcal{R}^4$ interactions are completely cancelled by two loop counterterms leaving no finite remainders as they would lead to terms inconsistent with perturbative string theory. The $D^6 \mathcal{R}^4$ interaction has $\Lambda^6$ divergence. This is cancelled by a counterterm which could leave a finite remainder determined by the genus two $D^6\mathcal{R}^4$ amplitude in string theory. However this finite remainder must vanish as a consequence of the structure of three loop supergravity~\cite{Basu:2014hsa}. This is because a five point two loop $\Lambda^6$ counterterm in the same supermultiplet is needed to cancel subleading divergences of the $D^6 \mathcal{R}^4$ interaction at three loops, and any finite remainder would be inconsistent with perturbative string theory.
For the $D^{12} \mathcal{R}^4$ interaction, the counterterm that cancels the ${\rm ln} (\Lambda l_{11})$ divergence can leave a finite remainder determined by the genus three amplitude of the $D^{12} \mathcal{R}^4$ interaction. The renormalized value of this counterterm will not be needed in our analysis.

All other two loop interactions at higher orders in the momentum expansion are finite. However, for the various interactions discussed above (as well as those in the same supermultiplet) which have a primitive two loop divergence, apart from the finite contributions, there are possible one loop subdivergences. Only the $\Lambda^3$ subdivergence yields a finite remainder using \C{defc}, which is true for the $D^4\mathcal{R}^4$, $D^6\mathcal{R}^4$ and $D^8\mathcal{R}^4$ interactions. For the $D^4\mathcal{R}^4$, $D^6\mathcal{R}^4$ and $D^8\mathcal{R}^4$ interactions, this leads to terms~\cite{Green:1999pu,Basu:2014hsa} 
\bea \label{1loopsubdiv}l_{11}^3 \int d^9 x \sqrt{-G^{(9)}} \mathcal{V}_2^{-3/2} D^4 \mathcal{R}^4 (\Lambda l_{11})^3 E_{5/2} (\Omega,\bar\Omega), \non \\  l_{11}^5 \int d^9 x \sqrt{-G^{(9)}} \mathcal{V}_2^{-1/2} D^6 \mathcal{R}^4 (\Lambda l_{11})^3 E_{3/2} (\Omega,\bar\Omega),\non \\ l_{11}^7 \int d^9 x \sqrt{-G^{(9)}} \mathcal{V}_2^{1/2} D^8 \mathcal{R}^4 (\Lambda l_{11})^3 E_{1/2} (\Omega,\bar\Omega)\eea      in the effective action respectively. The details for the $D^8\mathcal{R}^4$ interaction are given in appendix A leading to \C{Div4}. Thus, these lead to terms in the effective action of the type IIB theory given by
\bea \label{IR}&&l_s^3 \int d^9 x \sqrt{-g^{B}} r_B  D^4 \mathcal{R}^4 (\Lambda l_{11})^3  \Big( 2\zeta(5) e^{-2\phi^B} + \frac{8}{3}\zeta(4)e^{2\phi^B} +\ldots\Big), \non \\  &&l_s^5 \int d^9 x \sqrt{-g^{B}} r_B^{-1} D^6 \mathcal{R}^4 (\Lambda l_{11})^3 \Big(2\zeta(3) +4\zeta(2)e^{2\phi^B}+\ldots\Big),\non \\ &&2l_s^7 \int d^9 x \sqrt{-g^{B}} r_B^{-3} e^{2\phi^B} D^8 \mathcal{R}^4 (\Lambda l_{11})^3 {\rm ln}\Big(\frac{e^{-\phi^B}}{4\pi e^{-\gamma}}\Big)+\ldots,\eea
where we have dropped exponentially suppressed corrections. These lead to finite contributions using the counterterm in \C{defc}. 

Note that the various terms obtained in the type IIB effective action are in the string frame, and hence the perturbative contributions must involve powers of $e^{-2\phi^B}$. This is not the case for the logarithmic term in the $D^{8} \mathcal{R}^4$ interaction in \C{IR}. This is because the $D^8\mathcal{R}^4$ interaction in 9 dimensions has an infrared logarithmic divergence, which is captured by \C{IR}~\cite{Bern:1998ug,Green:2008bf}. The scale of this infrared divergent logarithm is moduli dependent in the supergravity calculation. Thus we see how the counterterm analysis of the ultraviolet divergences gives us information about infrared divergences in the theory. These infrared effects are also present in the finite part of the two loop $D^8\mathcal{R}^4$ interaction.

\section{The structure of three loop ultraviolet divergences}

The leading interaction at three loops~\cite{Bern:2007hh,Bern:2008pv,Bern:2010ue} is the $D^6\mathcal{R}^4$ interaction which has $\Lambda^{15}$ primitive divergence. This is cancelled by a three loop counterterm with coefficient $z$ which leaves a finite remainder fixed by the genus two $D^6\mathcal{R}^4$ amplitude
given by~\cite{Basu:2014hsa}
\be h(\Lambda l_{11})^{15} + z = 24\zeta(4),\ee
where $h$ is an irrelevant constant.
The other divergent interactions are the non--BPS $D^8\mathcal{R}^4, \ldots, D^{20} \mathcal{R}^4$ interactions which have $\Lambda^{13}, \ldots, \Lambda$ primitive divergences respectively. The primitive divergences of these interactions of the form $D^{2n} \mathcal{R}^4$ ($4 \leq n \leq 10$) yield the terms
\be l_{11}^{2n-1} \int d^9 x \sqrt{-G^{(9)}} \mathcal{V}_2 D^{2n} \mathcal{R}^4 (\Lambda l_{11})^{21-2n}\ee
in the effective action, leading to terms in the effective action
\be l_s^{2n-1} \int d^9 x \sqrt{-g^{B}} r_B^{-(1+2n/3)} e^{2n\phi^B/3} D^{2n} \mathcal{R}^4 (\Lambda l_{11})^{21-2n}\ee
in the type IIB theory. Consistency with perturbative string theory shows that all the primitive three loop divergences are cancelled by counterterms without leaving finite remainders, expect the ones for the $D^{12} \mathcal{R}^4$ and $D^{18} \mathcal{R}^4$ interactions which can receive finite contributions determined by the structure of the genus three $e^{4\phi^B} r_B^{-5}$ and genus four $e^{6\phi^B}r_B^{-7}$ amplitudes respectively. Hence we see the structure of the counterterms imposes strong constraints on the couplings.  

Each of these interactions also have subdivergences which must be cancelled by one and two loop counterterms. Among these subdivergences at one loop, only those of the form $\Lambda^3$ can leave finite remainders using \C{defc}, while the rest must vanish to be consistent with perturbative string theory. Two loop divergences which can leave a finite remainder must be of the form  $\Lambda^6$ or ${\rm ln}\Lambda$ as determined by the structure of the one and two loop counterterms. While the three loop ${\rm ln}\Lambda$ subdivergence must be cancelled by the two loop $D^{12} \mathcal{R}^4$ counterterm, the $\Lambda^6$ subdivergence must be cancelled by a product of two $\Lambda^3$ $\mathcal{R}^4$ counterterms.

All other interactions at higher orders in the momentum expansion are finite. Our aim is to analyze the constraints imposed by counterterms on the $D^8 \mathcal{R}^4$ interaction.
       
\section{The counterterm contributions to the $D^8\mathcal{R}^4$ interaction}

The three loop four graviton amplitude is given by~\cite{Bern:2007hh,Bern:2008pv,Bern:2010ue}
\bea \label{totcont}
\mathcal{A}_4^{(3)} = \frac{(4\pi^2)^3 \kappa_{11}^8}{(2\pi)^{33}}\sum_{S_3} \Big[ I^{(a)} + I^{(b)} + \frac{1}{2} I^{(c)} + \frac{1}{4} I^{(d)} + 2 I^{(e)} + 2 I^{(f)} + 4 I^{(g)} + \frac{1}{2} I^{(h)} + 2 I^{(i)}\Big] \mathcal{K}, \non \\ \eea
where $\mathcal{K}$ is the linearized approximation to $\mathcal{R}^4$ in momentum space, and $2\kappa_{11}^2 = (2\pi)^8 l_{11}^9$. Also $S, T$ and $U$ are the Mandelstam variables defined by $S = - G^{MN} (k_1+ k_2)_M (k_1 + k_2)_N, T = -G^{MN} (k_1+ k_4)_M(k_1 + k_4)_N$ and $U = - G^{MN} (k_1 + k_3)_M (k_1 + k_3)_N$, where $G_{MN}$ is the M theory metric, and the external momenta are labelled by $k_{iM} (i=1,\ldots,4)$ and satisfy $\sum_i k_{iM} =0$ and $k_i^2 =0$.
Also $S_3$ represents the 6 independent permutations of the external legs marked $\{1,2,3\}$ keeping the external leg $\{4\}$ fixed. The external momenta are directed inwards in all the loop diagrams in figures 1 and 2. We shall use the notation
\be \s_n = S^n + T^n + U^n \ee
throughout. We denote the low momentum expansion of the analytic part of the amplitude by
\bea \mathcal{A}_4^{(3)}= \frac{(4\pi^2)^3 \kappa_{11}^8}{(2\pi)^{33}} [I_3 + I_4 +\ldots], \eea
where $I_n$ receives contributions at $O(\s_n)$, and hence we are interested in $I_4$. 

The primitive $D^8\mathcal{R}^4$ $\Lambda^{13}$ divergence cancels as discussed before, and hence we need to consider the one and two loop subdivergences only.

\begin{figure}[ht]
\begin{center}
\[
\mbox{\begin{picture}(350,80)(0,0)
\includegraphics[scale=.55]{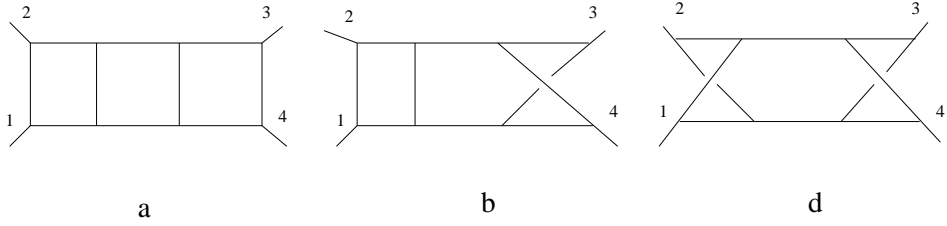}
\end{picture}}
\]
\caption{Three loop diagrams from the ladder skeleton}
\end{center}
\end{figure}

\begin{figure}[ht]
\begin{center}
\[
\mbox{\begin{picture}(350,200)(0,0)
\includegraphics[scale=.6]{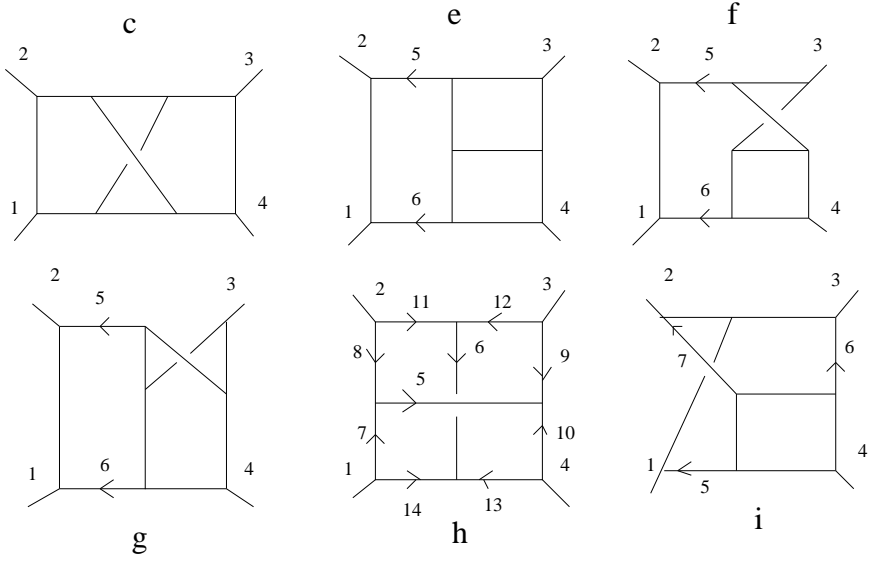}
\end{picture}}
\]
\caption{Three loop diagrams from the Mercedes skeleton}
\end{center}
\end{figure}

In \C{totcont}, the denominators for the various integrands in the loop diagrams are standard massless scalar field propagators, while the numerators $N^{(x)}$ are given by~\cite{Bern:2008pv}
\bea \label{num}
N^{(a)} &=& N^{(b)} = N^{(c)} = N^{(d)} = S^4 , \non \\ N^{(e)} &=& N^{(f)} = N^{(g)} = S^2 \tau_{35} \tau_{46}, \non \\ N^{(h)} &=& \Big(S(\tau_{26} +\tau_{36}) +T(\tau_{15} +\tau_{25}) +ST\Big)^2 \non \\ &&+ \Big(S^2 (\tau_{26} +\tau_{36}) - T^2 (\tau_{15} +\tau_{25}) \Big) \Big(\tau_{17} +\tau_{28} +\tau_{39} +\tau_{4,10} \Big) \non \\&& +S^2 (\tau_{17}  \tau_{28} +\tau_{39} \tau_{4,10}) +T^2 (\tau_{28} \tau_{39}+ \tau_{17} \tau_{4,10}) + U^2 (\tau_{17} \tau_{39} + \tau_{28} \tau_{4,10}), \non \\ N^{(i)} &=& (S\tau_{45} - T\tau_{46})^2 -\tau_{27} (S^2 \tau_{45} + T^2 \tau_{46}) - \tau_{15} (S^2 \tau_{47}+ U^2 \tau_{46})\non \\ &&-  \tau_{36} (T^2 \tau_{47} +U^2 \tau_{45}) - l_5^2 S^2 T - l_6^2 S T^2 +\frac{l_7^2}{3}STU,\eea
where
\be \tau_{ij} = -2 k_i \cdot l_j ~(i \leq 4, j \geq 5).\ee
The various momenta $l_i$ are denoted in figure 2.

In the calculations below, we drop various one and two loop counterterm contributions that do not leave any finite remainder as discussed above. Hence for the $D^8\mathcal{R}^4$ interaction we need to consider only the $O(\Lambda^3)$ and $O(\Lambda^6)$ divergences. 
Also while we refer to each contribution mentioned in \C{totcont} as $I^{(x)}$, the total contribution after the sum over $S_3$ is performed is referred to as $I^{(X)}$.

\subsection{Contributions from ladder skeleton loop diagrams}

These are contributions to $I_4$ from the loop diagrams $a,b$ and $d$ in figure 1. 

\subsubsection{$O(\Lambda^3)$ counterterm contributions}

We first consider the contribution from the diagram $a$. We give some of the details of the analysis for $I^{(a)}$ and for all the others, we simply write down the answer. For $I^{(a)}$, in the uncompactified theory from \C{num} we see that the one loop subdivergences are given by
\be \label{Ia}I^{(a)} = S^4 \int \frac{d^{11}r}{r^8} \Big[ 2 \int d^{11}p \int d^{11}q \frac{1}{p^6 q^4 (p+q)^2} + \int \frac{d^{11} p}{p^6} \int \frac{d^{11} q}{q^6}  \Big] .\ee 
These must be regularized by including the contributions from one loop counterterms. We now evaluate the integrals in \C{Ia} as well as the others to follow in the background $T^2 \times \mathbb{R}^{8,1}$. For all the cases, the 11 dimensional loop momentum $p_M$ decomposes as $\{p_\mu, l_I/l_{11}\}$ where $p_\mu$ is the 9 dimensional momentum and $l_I$ ($I=1,2$) is the KK momentum along $T^2$. Thus consider the integral
\be \label{Ia2}\int d^{11}p \int d^{11}q \frac{1}{p^6 q^4 (p+q)^2}\ee
in \C{Ia} in the compactified theory. We introduce a Schwinger parameter for every propagator. Hence the product of the propagators in the compactified theory coming from the denominator of \C{Ia2} is given by
\be \int_0^\infty \prod_{i=1}^6 d\s^i e^{-\sum_{j=1}^6 \s^j q_j^2} e^{-\Big((\s_1 +\s_2 +\s_3){\bf{m}}^2 +(\s_4 +\s_5) {\bf{n}}^2 + \s_6 {\bf{(m+n)}}^2\Big)/l_{11}^2},\ee     
where
\be q_j = \{p,p,p,q,q,p+q\}\ee
and
\be {\bf{m}}^2 \equiv G^{IJ} m_I m_J,\ee
where the metric of $T^2$ is given by
\be \label{defmet}
G_{IJ} = \frac{\mathcal{V}_2}{\Omega_2} \left( \begin{array}{cc} \vert \Omega \vert^2 & -\Omega_1 \\ -\Omega_1 & 1 \end{array} \right).\ee
Thus compactified on $T^2$, \C{Ia2} becomes
\be \frac{1}{(4\pi^2 l_{11}^2 \mathcal{V}_2)^2} \int_0^\infty \prod_{i=1}^6 d\s^i F_L(\s,\l,\r) \int d^9 p \int d^9 q e^{-\s p^2 - \l q^2 -\r (p+q)^2}\ee
where
\be F_L (\s,\lambda,\rho) = \sum_{m_I, n_I}e^{-G^{IJ} \Big( \s m_I m_J + \lambda n_I n_J +\rho(m+n)_I (m+n)_J\Big)/l_{11}^2},\ee
and we have defined
\be \s= \s_1 +\s_2 +\s_3, \quad \l = \s_4 +\s_5, \quad \r = \s_6.\ee
We now define\footnote{Thus
\be 0 \leq u_1, u_2, v_1 \leq 1, \quad u_1 \leq u_2.\ee} 
\be u_1 = \frac{\s_1}{\s}, \quad u_2 = \frac{\s_1 +\s_2}{\s}, \quad v_1 = \frac{\s_4}{\s},\ee
leading to
\be \int_0^\infty \prod_{i=1}^6 d\s^i= \int_0^\infty d\s d\l d\r \s^2 \l\int_0^1 d u_2 \int_0^{u_2} d u_1 \int_0^1 d v_1.\ee
Finally using
\be \int d^9 p \int d^9 q e^{-\s p^2 - \l q^2 -\r (p+q)^2} = \pi^{9/2} \Delta^{-9/2}\ee
where
\be \Delta = \s\lambda +\lambda\rho +\rho\s,\ee
\C{Ia2} when compactified on $T^2$ gives us
\be \label{I1}\mathcal{I}_1 = \frac{\pi^9}{(4\pi^2 l_{11}^2 \mathcal{V}_2)^2} \int_0^\infty d\s d\lambda d\rho \frac{\s^2\lambda}{2\Delta^{9/2}} F_L (\s,\lambda,\rho).\ee
All the integrals we need can be done using the same logic and so we only give the final answers. 
Hence in the 9 dimensional theory compactified on $T^2$, to cancel this $\Lambda^3$ divergence we must introduce the counterterm
\be \delta^{(3)} I^{(A)} = 2 \s_4 \cdot \frac{c_1 \pi^3}{12l_{11}^3 (4\pi^2)}\Big[ 2 \mathcal{I}_1+ \mathcal{I}_3\Big], \ee 
where
\be \label{I3} \mathcal{I}_3 = \frac{\pi^{11}}{4(4\pi^2)^2 l_{11}^{10}} \Big(\frac{2}{5} (\Lambda l_{11})^5 + \frac{3}{4\pi^{9/2}} \mathcal{V}_2^{-5/2} E_{5/2} (\Omega,\bar\Omega)\Big)^2.\ee
The counterterm involving $\mathcal{I}_1$ is depicted by $a$ in figure 3, while the one involving $\mathcal{I}_3$ is depicted in figure 4. In the integral $\mathcal{I}_3$ which involves the one loop integral, we have performed Poisson resummation to go from the KK momenta to winding momenta to perform the integrals.  

\begin{figure}[ht]
\begin{center}
\[
\mbox{\begin{picture}(280,80)(0,0)
\includegraphics[scale=.6]{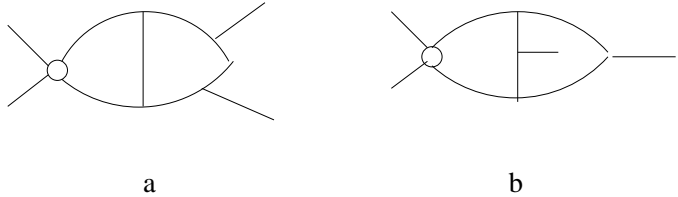}
\end{picture}}
\]
\caption{Planar and non--planar one loop counterterms for the ladder skeleton diagrams}
\end{center}
\end{figure}

\begin{figure}[ht]
\begin{center}
\[
\mbox{\begin{picture}(220,90)(0,0)
\includegraphics[scale=.35]{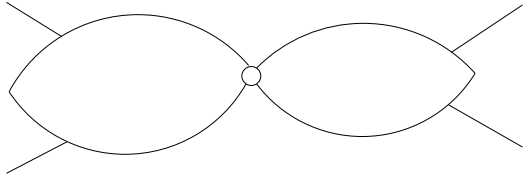}
\end{picture}}
\]
\caption{Another one loop counterterm for the ladder skeleton diagrams}
\end{center}
\end{figure}

Similarly from figure $b$, divergence cancellation gives the counterterm
\be \delta^{(3)} I^{(B)} = 2\s_4 \cdot \frac{c_1 \pi^3}{12l_{11}^3 (4\pi^2)} \Big[ \mathcal{I}_1+ \mathcal{I}_2\Big]\ee
in the compactified theory, where
\be \label{I2}\mathcal{I}_2 = \frac{\pi^9}{(4\pi^2 l_{11}^2 \mathcal{V}_2)^2} \int_0^\infty d\s d\lambda d\rho \frac{\s\lambda\rho}{\Delta^{9/2}} F_L (\s,\lambda,\rho).\ee
This counterterm is depicted by $b$ in figure 3.

Finally from $I^{(d)}$, we get the counterterm 
\be \delta^{(3)} I^{(D)} = 2\s_4 \cdot \frac{c_1 \pi^3}{12l_{11}^3 (4\pi^2)} \cdot 2 \mathcal{I}_2\ee
in the compactified theory.

Thus in \C{totcont} to cancel the $\Lambda^3$ divergence we get the total counterterm contribution 
\be \label{add} \delta^{(3)} I^{(A)} + \delta^{(3)}I^{(B)} +\frac{1}{4} \delta^{(3)}I^{(D)} = 2\s_4  \cdot \frac{c_1 \pi^3}{12l_{11}^3 (4\pi^2)}\Big[ \mathcal{I}_{12}+\mathcal{I}_3\Big],\ee
where
\be \label{add2}\mathcal{I}_{12} = \frac{\pi^9}{4(4\pi^2 l_{11}^2 \mathcal{V}_2)^2} \int_0^\infty d\s d\lambda d\rho \Delta^{-9/2} \Big[ \Delta (\s+\lambda+\rho) + 3\s\lambda\rho\Big] F_L (\s,\lambda,\rho).\ee

Apart from cancelling the divergence, from the term involving $\mathcal{I}_3$ in \C{add}, on using \C{defc} we get a finite contribution to $I_4$ in the regularized theory given by
\be I_4 = \frac{\s_4}{(4\pi^2)^3l_{11}^{13}} \cdot \frac{\pi^7}{64} \mathcal{V}_2^{-5} E^2_{5/2} (\Omega,\bar\Omega).\ee 
This leads to a term in the effective action of the form
\bea \label{D8R4}l_{11}^7 \int d^9 x \sqrt{-G^{(9)}} \mathcal{V}_2^{-4}  D^8 \mathcal{R}^4 E_{5/2}^2 (\Omega,\bar\Omega),\eea
which in the type IIB theory becomes
\be \label{inconsistent}l_s^7 \int d^9 x \sqrt{-g^{B}} r_B^3 \Big(4\zeta(5)^2 e^{-4\phi^B} +\frac{32}{3} \zeta(4)\zeta(5) + \frac{64}{9} \zeta(4)^2 e^{4\phi^B}\Big)D^8 \mathcal{R}^4 +\ldots.\ee
The first term is inconsistent with string perturbation theory, and hence this must cancel when other contributions of this kind are added. The second and third terms yield contributions at genus one and three respectively. Contributions of this type generalize for higher derivative interactions as described in appendix C for the $D^{10} \mathcal{R}^4, D^{12} \mathcal{R}^4, D^{14} \mathcal{R}^4$ and $D^{16} \mathcal{R}^4$ interactions.

In fact, the genus one contribution in \C{inconsistent} gives 
\be \mathcal{A}_4^{(3)} = (2\pi^8 l_{11}^{15} r_B) \mathcal{K} r_B (\s_4 l_s^8) \frac{\pi^2 \zeta(5) r_B^2}{2 \cdot4!6!}\ee
to the amplitude. The overall factor of $2\pi^8 l_{11}^{15} r_B$ is needed to obtain the correct normalization and is common to the multiloop amplitudes. The remaining part yields a contribution of the form $\pi^2\zeta(5) r_B^3$ at genus one for the $D^8\mathcal{R}^4$ amplitude. In fact, this structure precisely agrees with that obtained from string perturbation theory~\cite{Green:2008uj}\footnote{One has to send $r_B \rightarrow r_B^{-1}$ to see the agreement using the pertubative equality of the type IIA and IIB amplitudes.} upto a numerical factor. The overall factor is not expected to match as there are other contributions of this type as well. In fact, we shall encounter one such source of contribution later on.

We shall consider the finite contribution to $I_4$ coming from the $\mathcal{I}_{12}$ term in \C{add} later. 

\subsubsection{$O(\Lambda^6)$ counterterm contributions}

We now consider the $\Lambda^6$ divergences. From \C{num},
we get that
\be I^{(A)} = I^{(B)} = I^{(D)} ,\ee
leading to
\be \label{totoneloop}\delta^{(3)}I^{(A)} + \delta^{(3)}I^{(B)} +\frac{1}{4} \delta^{(3)}I^{(D)} =\frac{\s_4}{(4\pi^2)^3 l_{11}^{13}}\cdot \frac{9\pi^{11/2}}{2} \cdot \Big( \frac{c_1 \pi^3}{12} \Big)^2.\Big[ \frac{2}{7} (\Lambda l_{11})^7 +\frac{15}{8\pi^{13/2}}\mathcal{V}_2^{-7/2} E_{7/2} (\Omega,\bar\Omega)\Big].\ee

This counterterm is depicted in figure 5.

\begin{figure}[ht]
\begin{center}
\[
\mbox{\begin{picture}(220,60)(0,0)
\includegraphics[scale=.5]{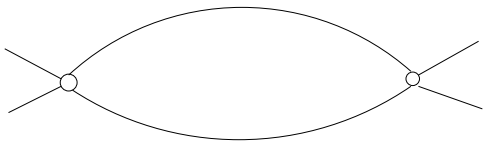}
\end{picture}}
\]
\caption{Product of one loop counterterms for the ladder skeleton diagrams}
\end{center}
\end{figure}

Thus from \C{defc}, we get a finite contribution to $I_4$ given by
\be I_4 = \frac{\s_4}{(4\pi^2)^3 l_{11}^{13}} \cdot \frac{5\pi^9}{8 \cdot 24} \mathcal{V}_2^{-7/2} E_{7/2} (\Omega,\bar\Omega).\ee
This leads to a term in the effective action of the form
\bea \label{finite}&&l_{11}^7 \int d^9 x \sqrt{-G^{(9)}} \mathcal{V}_2^{-5/2}  D^8 \mathcal{R}^4 E_{7/2} (\Omega,\bar\Omega)\non \\ &&= l_{11}^7 \int d^9 x \sqrt{-G^{(9)}} \mathcal{V}_2^{-5/2}  D^8 \mathcal{R}^4 \Big( 2\zeta(7)\Omega_2^{7/2} +\frac{32}{15} \zeta (6) \Omega_2^{-5/2}+\ldots\Big),\eea
which in the type IIB theory becomes
\be  l_s^7 \int d^9 x \sqrt{-g^{B}} r_B \Big( 2\zeta(7)e^{-2\phi^B} +\frac{32}{15} \zeta (6) e^{4\phi^B} +\ldots\Big) D^8 \mathcal{R}^4 .\ee
This leads to terms involving genus zero and genus three in the type IIB theory\footnote{Such counterterm contributions to the $D^{10} \mathcal{R}^4, D^{12} \mathcal{R}^4$ and $D^{14} \mathcal{R}^4$ interactions have been discussed in appendix C.}. However, we see that this does not yield the complete genus zero amplitude which has a different coefficient. This is not a contradiction as we have only considered the contributions that arise from counterterms at three loops. We have not considered the finite contributions as well as contributions from higher loops. In fact, it is not difficult to see that there are indeed contributions of the kind \C{finite} that arise at four loops. Consider the four loop diagram $a$ in figure 6 which has the $D^8\mathcal{R}^4$ interaction as the leading term in the low momentum expansion~\cite{Bjornsson:2010wm,Bjornsson:2010wu}. The primitive four loop $D^8\mathcal{R}^4$ divergence is $\Lambda^{22}$. There is a subleading three loop $\Lambda^{15}$ divergence which is cancelled by the three loop five point counterterm $b$ in figure 6. Thus at four loops, this counterterm yields a contribution of the form 
\be \delta^{(4)}I^{(4)} \sim \frac{\s_4 \hat{z}}{l_{11}^{22}} \Big[ \frac{2}{7} (\Lambda l_{11})^7 +\frac{15}{8\pi^{13/2}}\mathcal{V}_2^{-7/2} E_{7/2} (\Omega,\bar\Omega)\Big],\ee 
where the analytic part of the four loop four graviton amplitude is expanded as  
\be \mathcal{A}_4^{(4)} = \frac{(4\pi^2)^4\kappa_{11}^{10}}{(2\pi)^{44}} \Big[ I^{(4)} +\ldots\Big].\ee
Here the three loop five point counterterm which has coefficient proportional to $\hat{z}$ lies in the same supermultiplet as the three loop primitive counterterm for the $D^6\mathcal{R}^4$ interaction and hence~\cite{Basu:2014hsa}
\be \hat{z} \sim (\Lambda l_{11})^{15} +\zeta(4).\ee  
It immediately follows that we get a finite contribution of the form \C{finite} in the effective action. While there are several contributions of this kind, we see that simply the three loop counterterm analysis shows the existence of this term in the effective action with the correct perturbative structure and transcendentality of the various coefficients.

\begin{figure}[ht]
\begin{center}
\[
\mbox{\begin{picture}(280,150)(0,0)
\includegraphics[scale=.50]{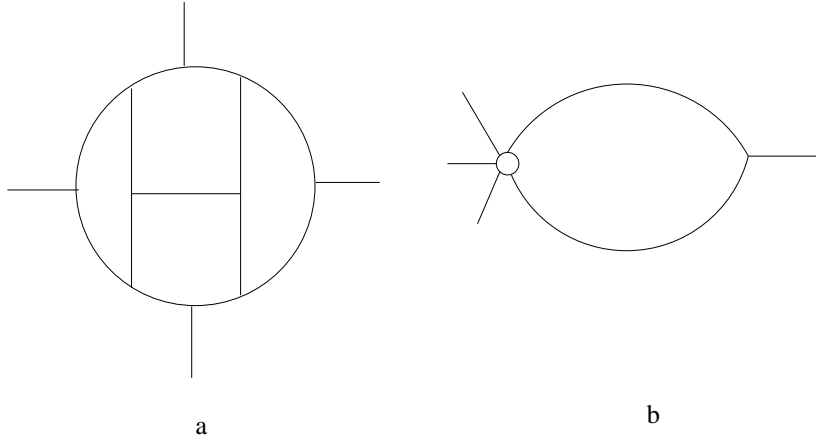}
\end{picture}}
\]
\caption{A four loop diagram and a three loop counterterm}
\end{center}
\end{figure}

\subsection{Contribution from Mercedes skeleton loop diagrams}

These are contributions from loop diagrams $c,e,f,g,h$ and $i$ in figure 2.

\subsubsection{$O(\Lambda^3)$ counterterm contributions}

There are no contributions from $I^{(c)}$.

From the loop diagram diagram $e$, we get $\Lambda^3$ divergent contributions which have to be cancelled by one loop counterterms. We describe this case in detail, as the analysis for the other cases proceeds along the same lines. 

For $I^{(e)}$, from \C{num} we see that the one loop subdivergences are given by
\bea \label{fige}I^{(e)}= -4 S^2 \int\frac{d^{11}r}{r^8} (\mathcal{J}_1 +\mathcal{J}_2) - 4S^2 \int \frac{d^{11}r (k_3 \cdot r)(k_4 \cdot r)}{r^{10}}\mathcal{J}_3\eea
where
\bea \label{defJ}
\mathcal{J}_1 &=& \int \frac{d^{11}p d^{11}qk_3 \cdot (q+k_1 + k_2) k_4 \cdot q}{q^2 (q+k_1)^2(q+k_1 + k_2)^2 p^2 (p+k_4)^2(p+q)^2}, \non \\ \mathcal{J}_2 &=& \int \frac{d^{11}p d^{11}qk_3 \cdot (q+k_1 + k_2) k_4 \cdot q}{q^2 (q+k_1)^2(q+k_1 + k_2)^2 p^2 (p+k_3)^2(p-q+ k_3 + k_4)^2}, \eea
and
\bea \mathcal{J}_3 &=& \int \frac{d^{11}pd^{11}q}{p^2 (p+k_4)^2 q^2 (q+k_3)^2(p+q+k_3 + k_4)^2},\eea
and the $O(k^8)$ terms have to be kept in $I^{(e)}$. The divergences in the terms involving $\mathcal{J}_1$ and $\mathcal{J}_2$ in \C{fige} are cancelled by a four point counterterm depicted by $a$ in figure 7, while the divergences in the term involving $\mathcal{J}_3$ is cancelled by a five point counterterm in figure 8. 

Thus the counterterm is given by
\be \delta^{(3)} I^{(e)} = -4S^2 \cdot \frac{c_1 \pi^3}{12 l_{11}^3 (4\pi^2)}(\mathcal{J}_1 +\mathcal{J}_2) - 4S^3 \cdot \frac{\hat{c}_1 \pi^3}{l_{11}^3 (4\pi^2)}\mathcal{J}_3,\ee
where $\hat{c}_1$ is the coefficient of the one loop five point counterterm which is is the same supermultiplet as the $\mathcal{R}^4$ counterterm.

The expressions for $\mathcal{J}_1$ and $\mathcal{J}_3$ have been obtained in \C{J1} and \C{J3} respectively, while $\mathcal{J}_2$ can be calculated similarly. 

\begin{figure}[ht]
\begin{center}
\[
\mbox{\begin{picture}(220,130)(0,0)
\includegraphics[scale=.6]{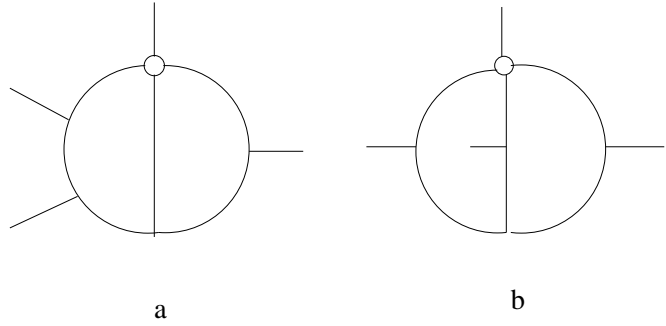}
\end{picture}}
\]
\caption{The planar and non--planar one loop four point counterterm diagrams}
\end{center}
\end{figure}

\begin{figure}[ht]
\begin{center}
\[
\mbox{\begin{picture}(100,90)(0,0)
\includegraphics[scale=.6]{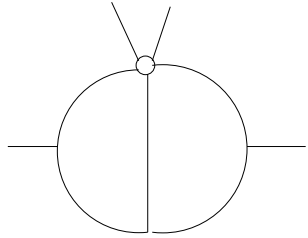}
\end{picture}}
\]
\caption{The one loop five point counterterm diagram}
\end{center}
\end{figure}

One can calculate all the contributions coming from all the other loop diagrams in exactly the same way as described in appendix B. However, we shall refrain from doing so, as the calculations are quite cumbersome. Instead we shall consider the constraints imposed by the symmetries of the two loop skeleton diagram on the structure of the integrals, which will turn out to be enough for our purposes. It is easy to see what are the one loop counterterm diagrams that contribute at $O(\Lambda^3)$. Diagram $f$ only involves $a$ in figure 7, while $g$ involves $a$ in figure 7 and figure 8. Diagram $h$ only involves figure 8, and $i$ involves figure 8 and $b$ in figure 7. Ignoring the various numerical factors, we now consider the various integrals that arise while calculating the various diagrams. Though this tedious exercise involves considering the contributions from each diagram separately and including the numerators in \C{num}, the analysis proceeds in a reasonably straightforward way. The final complete $\Lambda^3$ counterterm contribution to $I_4$ is of the form
\be \label{finalct}
\delta^{(3)} I_4 = \frac{c_1 \pi^3\s_4}{l_{11}^3(4\pi^2)} \cdot \frac{\pi^9}{( 4\pi^2 l_{11}^2 \mathcal{V}_2)^2} \int_0^\infty d\s d\l d\r F_L (\s,\l,\r)\mathcal{P} (\s,\l,\r)\ee     
where 
\bea \label{P} \mathcal{P} (\s,\l,\r) &=& \frac{\hat{a}_1 \s^2 \l + \hat{a}_2\s\l\r}{\Delta^{9/2}} + \frac{\hat{a}_3\s^3\l^2 + \hat{a}_4\s^3 \l\r + \hat{a}_5 \s^2\l^2\r}{\Delta^{11/2}} \non \\ &&+\frac{\hat{a}_6\s^4 \l^3 + \hat{a}_7 \s^4 \l^2 \r+ \hat{a}_8 \s^3\l^3\r+ \hat{a}_9\s^3\l^2\r^2}{\Delta^{13/2}},\non \\ \eea
and we have not distinguished between $c_1$ and $\hat{c}_1$ in the coefficient of \C{finalct} because both leave a finite remainder proportional to $\zeta (2)$ using \C{defc}. In \C{P}, the various constants $\hat{a}_i$ have vanishing transcendentality. 

We now analyze the structure that arises in \C{P}. The series terminates at $O(\Delta^{-13/2})$ because there are at most two derivatives of Schwinger parameters acting on $\Delta^{-9/2}$. At every fixed order in $\Delta$, the number of Schwinger parameters is completely determined by dimensional analysis as each Schwinger parameter has dimension $l_{11}^2$. In fact, there are several vanishing terms in \C{P}:

(i) $\s^3$ term vanishes in the $\Delta^{-9/2}$ term,

(ii) $\s^5$, $\s^4 \l$ terms vanish in the $\Delta^{-11/2}$ term,

(iii) $\s^7,$ $\s^6 \l$, $\s^5 \l^2$, $\s^5 \l\r$ terms vanish in the $\Delta^{-13/2}$ term.

In \C{P}, at every order in $\Delta$, we have not written the terms that are related by symmetries of the two loop skeleton diagram\footnote{The symmetry group of the two loop skeleton diagram is $S_3$, which amounts to interchanging the three Schwinger parameters among themselves.} and each term is independent. While most of the terms that vanish do not arise at all, the vanishing of the $\s^5 \l^2$ and $\s^5\l\r$ terms in the $\Delta^{-13/2}$ term happens from cancellations between various contributions to the integrand. A possible $\s^5 \l^2$ term always arises only in the combination (in arbitrary dimensions $D$) 
\bea \label{van1}&&\int_0^\infty d\s d\l d\r F_L \s^3 \l^2 \Big(\frac{\p^2}{\p\l^2} +\frac{\p^2}{\p\r^2} - 2\frac{\p^2}{\p\l \p\r}\Big) \Delta^{-D/2} \non \\ &&= \int_0^\infty d\s d\l d\r F_L \s^3 \l^2 \Big[ \frac{D(D+2)}{4\Delta^{D/2+2}} \Big( (\s+\r)^2 +(\s+\l)^2 -2(\s+\r)(\s+\l)\Big) +\frac{D}{\Delta^{D/2+1}}\Big],\non \\ \eea
and thus the $\s^5 \l^2$ term vanishes. The remaining terms are already in \C{P}. Similarly, the vanishing of the $\s^5 \l\r$ term happens because it always arises only in the combination 
\be \label{van2}\int_0^\infty d\s d\l d\r F_L \s^3 \l\r \Big(\frac{\p^2}{\p\l^2} +\frac{\p^2}{\p\r^2} - 2\frac{\p^2}{\p\l \p\r}\Big) \Delta^{-D/2}.\ee
Note that vanishing happens in all dimensions. The contributions of the type \C{van1} and \C{van2} only come from diagrams $f$ and $i$ respectively. 
 
We now express $\mathcal{P}(\s,\l,\r)$ in a manifestly $S_3$ invariant way. To do so, we write the various non--invariant expressions in \C{P} in terms of $S_3$ invariants. We construct the following nine invariants:

(i) At cubic order: 
\be p_1 = \s\l\r , \quad p_2 = \Delta (\s+\l+\r).\ee 
 
(ii) At fifth order:

\be p_3 = \Delta p_1 , \quad p_4 =\Delta p_2, \quad p_5 = \s\l\r(\s^2+\l^2+\r^2).\ee

(iii) At seventh order: 

\bea p_6 = \Delta p_3 , \quad p_7 = \Delta p_4,  \quad p_8 = \Delta p_5, \quad p_9 =(\s\l\r)^2 (\s+\l+\r).\eea

Thus insider the integral in \C{finalct}, we can make the replacements 

(i) at cubic order   
\bea \s\l\r \rightarrow p_1, \quad \s^2\l \rightarrow \frac{p_2}{6}  - \frac{p_1}{2} , \eea 

(ii) at fifth order 
\bea \s^3\l^2 \rightarrow \frac{p_4}{6} -\frac{p_5}{3} -\frac{5p_3}{6}, \quad \s^3\l\r \rightarrow \frac{p_5}{3}, \quad \s^2\l^2\r \rightarrow \frac{p_3}{3},\eea

(iii) at seventh order
\bea &&\s^4\l^3 \rightarrow \frac{p_7}{6} -\frac{p_8}{2} -\frac{7p_6}{6} +\frac{5p_9}{6}, \quad \s^4 \l^2\r \rightarrow \frac{p_8}{6} -\frac{p_9}{6}, \non \\ && \s^3\l^3\r \rightarrow \frac{p_6}{3} -\frac{2 p_9}{3}, \quad \s^3\l^2\r^2 \rightarrow \frac{p_9}{3}.\eea

Thus, we have that
\be \mathcal{P} (\s,\l,\r)= a_1 \frac{\s+\l+\r}{\Delta^{7/2}}+a_2\frac{\s\l\r}{\Delta^{9/2}}+a_3\frac{\s\l\r(\s^2+\l^2+\r^2)}{\Delta^{11/2}}+a_4\frac{(\s\l\r)^2(\s+\l+\r)}{\Delta^{13/2}},\ee
where the undetermined numerical factors $a_i$ have vanishing transcendentality.
Thus from \C{finalct}, we get that
\bea &&\delta^{(3)} I_4 = \frac{c_1\pi^{14} \s_4}{(4\pi^2)^3} \sum_{\hat{m}_I,\hat{n}_I} \int_0^\infty d\hat{\s} d\hat{\lambda} d{\hat\r} e^{-\pi^2 l_{11}^2 G_{IJ} \Big( \hat\l \hat{m}_I \hat{m}_J + \hat\s \hat{n}_I \hat{n}_J +\hat\r (\hat{m}+\hat{n})_I (\hat{m}+\hat{n})_J\Big)} \non \\ &&\times \Big[ a_1 (\hat\s+\hat\l+\hat\r)\hat\Delta^{1/2}+a_2\frac{\hat\s\hat\l\hat\r}{\hat\Delta^{1/2}}+a_3\frac{\hat\s\hat\l\hat\r(\hat\s^2+\hat\l^2+\hat\r^2)}{\hat\Delta^{3/2}}+a_4\frac{(\hat\s\hat\l\hat\r)^2(\hat\s+\hat\l+\hat\r)}{\hat\Delta^{5/2}}\Big] \non \\ \eea
where we have Poisson resummed to go from momentum modes to winding modes and defined~\cite{Green:1999pu,Green:2005ba}
\be \hat\rho = \frac{\rho}{\Delta}, \quad \hat\s = \frac{\s}{\Delta}, \quad \hat\lambda = \frac{\lambda}{\Delta},\ee
and
\be \hat\Delta = \hat\s\hat\rho +\hat\s\hat\lambda + \hat\rho\hat\lambda= \Delta^{-1}.\ee 
Thus $\hat\s,\hat\l,\hat\r$ have dimensions $l_{11}^{-2}$. 
Further defining
\be \tau_1 = \frac{\hat\rho}{\hat\rho +\hat\lambda}, \quad \tau_2 = \frac{\sqrt{\hat\Delta}}{\hat\rho+\hat\lambda}, \quad V_2 = l_{11}^2 \sqrt{\hat\Delta},\ee
we get that
\bea \label{expand}\delta^{(3)} I_4 = \frac{6c_1 \pi^{14}\s_4}{(4\pi^2)^3 l_{11}^{13}}\sum_{\hat{m}_I,\hat{n}_I}\int_0^\infty dV_2 V_2^4 \int_{\mathcal{F}} \frac{d^2\tau}{\tau_2^2} e^{-\pi^2 G_{IJ} (\hat{m} +\hat{n}\tau)_I (\hat{m}+\hat{n} \bar\tau)_J V_2/\tau_2} \mathcal{A} (\tau,\bar\tau)\eea
where $d^2 \tau= d\tau_1 d\tau_2$ and $\mathcal{F}$ is the fundamental domain of $SL(2,\mathbb{Z})$ defined by
\be \mathcal{F} = \{ -\frac{1}{2} \leq \tau_1 \leq \frac{1}{2}, \tau_2 \geq 0, \vert \tau \vert^2 \geq 1\}.\ee
Also $\mathcal{A}$ is given by
\bea \label{defA}\mathcal{A} (\tau,\bar\tau) &=& \frac{a_1}{\tau_2} (\tau_2^2 -T +1)+\frac{a_2 T}{\tau_2^3}(\tau_2^2 -T) +\frac{a_3 T}{\tau_2^5} (\tau_2^2 -T)\Big( 1-2T+ (\tau_2^2 -T)^2 \Big)\non \\&& +\frac{a_4 T^2}{\tau_2^7} (\tau_2^2 -T)^2 (\tau_2^2 - T+1),\eea
where
\be \label{defT}T = \vert \tau_1 \vert - \tau_1^2.\ee 
Thus the amplitude boils down to an integral over the moduli space of an auxiliary $T^2$ parametrized by volume $V_2$ and complex structure $\tau$, where the integral of the complex structure $\tau$ is over $\mathcal{F}$. The integrand involves an $SL(2,\mathbb{Z})$ invariant lattice factor, and a non--$SL(2,\mathbb{Z})$ invariant function $\mathcal{A}(\tau,\bar\tau)$.

Now let us consider the structure of \C{expand}. The integral has a leading ultraviolet divergence coming from the zero winding sector with $\hat{m}_I = \hat{n}_I =0$ in the lattice sum which is of the form $(\Lambda l_{11})^{10}$ arising from the boundary of the $V_2$ integral cutoff at $V_2 \sim (\Lambda l_{11})^2$. Along with \C{defc} this yields the two loop $\Lambda^{13}$ primitive divergence.  

The subleading divergence is calculated as in appendix A, leading to
\be \Delta_\Omega \delta^{(3)} I^{div}_4 = \frac{6 c_1 \pi^{14}\s_4}{(4\pi^2)^3 l_{11}^{13}} \int_0^\infty dV_2 V_2^4 \int_{-1/2}^{1/2} d\tau_1 \Big[ \mathcal{A} \frac{\p \hat{F}_L}{\p\tau_2} - \hat{F}_L \frac{\p\mathcal{A}}{\p\tau_2}\Big]\Big\vert_{\tau_2 = (\Lambda l_{11})^2/V_2} \ee
where $\hat{F}_L$ is defined in \C{deFF}. From \C{defA}, the relevant terms are
\be \mathcal{A} \rightarrow (a_1 +a_3 T)\tau_2, \quad \frac{\p\mathcal{A}}{\p\tau_2} \rightarrow (a_1 +a_3 T),\ee  
leading to
\be  \Delta_\Omega \delta^{(3)} I^{div}_4 = \frac{27 c_1 \pi^{19/2}\s_4}{(4\pi^2)^3 l_{11}^{13}} \Big( a_1 +\frac{a_3}{12} \Big) (\Lambda l_{11})^5 \mathcal{V}_2^{-5/2} E_{5/2} (\Omega,\bar\Omega),\ee
and thus
\be \delta^{(3)} I^{div}_4 \sim \Lambda^8 l_{11}^{-5} \Big( a_1 +\frac{a_3}{12} \Big) \s_4 \mathcal{V}_2^{-5/2} E_{5/2} (\Omega,\bar\Omega),\ee
on using \C{defc}. 
Hence the primitive and subleading divergences of the integral in \C{expand} give us that
\be \delta^{(3)} I_4 \sim \s_4 \Lambda^{13} + \s_4 \Lambda^8 l_{11}^{-5} \mathcal{V}_2^{-5/2} E_{5/2} (\Omega,\bar\Omega),\ee   
which cancel divergences in the three loop amplitude and leave no finite remainder. The finite remainder comes from the finite part of the integral in \C{expand} which we now consider.    

For this purpose, it is very useful to note that $\mathcal{A}$ splits into a sum of functions $\mathcal{A}_i$ each of which satisfies Poisson equation on $\mathcal{F}$. The structure of these equations is determined recursively~\cite{Green:2008bf}. We first start with the leading term in the small $\tau_2$ limit in  \C{defA}, which is 
\be \label{lead} \frac{a_4 T^4 (1-T)}{\tau_2^7},\ee  
and construct a Poisson equation which has \C{lead} as the dominant term in the small $\tau_2$ limit which we call $\mathcal{A}_1$ below. The leading subdominant term in this equation is $O(\tau_2^{-5})$, which is subtracted from the $O(\tau_2^{-5})$ terms in \C{defA} which yields the leading term in the next Poisson equation. This procedure proceeds recursively till all the terms are exhausted. We use the relations
\be \Big( \frac{\p T}{\p\tau_1}\Big)^2 =1-4T, \quad \frac{\p^2 T}{\p\tau_1^2} = 2(\delta(\tau_1)-1)\ee
repeatedly in our analysis. Thus we get that
\be \label{split}\mathcal{A} (\tau,\bar\tau)= \sum_{i=1}^6 \mathcal{A}_i (\tau,\bar\tau)\ee
where we now discuss the structure of $\mathcal{A}_i$.
Including the leading order term in the small $\tau_2$ expansion in the definition of $\mathcal{A}_1$, we get that
\be \mathcal{A}_1 = a_4 \hat{\mathcal{A}}_1,\ee
where $\hat{\mathcal{A}}_1$ satisfies the Poisson equation
\be \label{defa1} \Delta_\tau \hat{\mathcal{A}}_1 = 56 \hat{\mathcal{A}}_1 -\frac{980}{429} \tau_2 \delta(\tau_1) - \frac{150}{143} \tau_2\Big[ \tau_2^2+ \frac{1}{\tau_2^2}\Big]\delta(\tau_1).\ee
In \C{defa1}, $\hat{\mathcal{A}}_1$ is given by 
\bea \hat{\mathcal{A}}_1 &=& \frac{T^4 (1-T)}{\tau_2^7} +\frac{T^2 (6-38 T + 45T^2)}{13\tau_2^5} +\frac{3(1-25T +140T^2 -210T^3)}{143 \tau_2^3} \non \\ &&+\frac{5(11- 98 T+210 T^2)}{429 \tau_2} +\frac{5(11-45T)\tau_2}{429} +\frac{3 \tau_2^3}{143}.\eea
Note that we get an $O(\tau_2^3)$ term which is not there in \C{defA}, the total coefficient of which has to cancel among various $\mathcal{A}_i$. At the first subleading order we get that
\be \mathcal{A}_2 = -\Big(a_3 +\frac{6a_4}{13} \Big)\hat{\mathcal{A}}_2\ee
where
\be \Delta_\tau \hat{\mathcal{A}}_2 = 30 \hat{\mathcal{A}}_2 -\frac{100}{21} \tau_2 \delta(\tau_1) - \frac{8}{3} \tau_2\Big[ \tau_2^2+ \frac{1}{\tau_2^2}\Big]\delta(\tau_1).\ee
Again $\hat{\mathcal{A}}_2$ is given by
\be \hat{\mathcal{A}}_2 = \frac{T^2 (1-T)^2}{\tau_2^5} +\frac{1-12T +36T^2 -28T^3}{9\tau_2^3}+\frac{2(4-25T+35T^2)}{21\tau_2} +\frac{4(2-7T)\tau_2}{21} +\frac{\tau_2^3}{9}.\ee

The remaining Poisson equations for $\mathcal{A}_i$ exactly follow the same pattern. We do not write them down explicitly as they are quite messy, and simply write down their general form. The equations for $\mathcal{A}_3, \ldots, \mathcal{A}_6$ are given by 
\bea    && \Delta_\tau \mathcal{A}_3 = 12 \mathcal{A}_3 +(\tau_2^3 +\tau_2 +\tau_2^{-1})\delta(\tau_1), \non \\  && \Delta_\tau \mathcal{A}_4 = 2 \mathcal{A}_4 +(\tau_2^3 +\tau_2)\delta(\tau_1), \non \\  && \Delta_\tau \mathcal{A}_5 =  \tau_2^3 \delta(\tau_1), \non \\  && \Delta_\tau \mathcal{A}_6 = 6 \mathcal{A}_6,\eea
where $\mathcal{A}_i$ takes the form
\bea  &&\mathcal{A}_3 \sim \frac{1+T+T^2 +T^3}{\tau_2^3} +\frac{1+T+T^2}{\tau_2} +\tau_2 (1+T) +\tau_2^3, \non \\ &&\mathcal{A}_4 \sim \frac{1+T+T^2}{\tau_2} +\tau_2 (1+T) +\tau_2^3, \non \\ &&\mathcal{A}_5 \sim \tau_2 (1+T) +\tau_2^3, \non \\ &&\mathcal{A}_6 \sim \tau_2^3,\eea
where the various neglected coefficients are linear combinations of $a_i$. Note that the eigenvalue in every Poisson equation for $\mathcal{A}_i$ is easily determined by the power of $\tau_2$ in the leading contribution for small $\tau_2$.  

Now the finite part of \C{expand} comes from the lattice sum where
\be \label{bdyvan}(\hat{m}_1,\hat{m}_2) \neq (0,0), (\hat{n}_1,\hat{n}_2) \neq (0,0).\ee 
To obtain this contribution, we use the relation
\be \Delta_\Omega \sum_{\hat{m}_I,\hat{n}_I }e^{-\pi^2 G_{IJ} (\hat{m} +\hat{n}\tau)_I (\hat{m}+\hat{n} \bar\tau)_J V_2/\tau_2}= \Delta_\tau \sum_{\hat{m}_I,\hat{n}_I }e^{-\pi^2 G_{IJ} (\hat{m} +\hat{n}\tau)_I (\hat{m}+\hat{n} \bar\tau)_J V_2/\tau_2}.\ee
Thus defining
\be \delta^{(3)} I_4^{finite} = \frac{6c_1 \pi^{14}\s_4}{(4\pi^2)^3l_{11}^{13}}\sum_{i=1}^6 I_i^{finite}\ee
we get that
\be \label{poissonexp}\Delta_\Omega I_i^{finite} = \sum' \int_0^\infty dV_2 V_2^4 \int_{\mathcal{F}} \frac{d^2\tau}{\tau_2^2}e^{-\pi^2 G_{IJ} (\hat{m} +\hat{n}\tau)_I (\hat{m}+\hat{n} \bar\tau)_J V_2/\tau_2} \Delta_\tau \mathcal{A}_i\ee
where we have integrated by parts twice, and the boundary contributions vanish using \C{bdyvan}. The sum in \C{poissonexp} stands for the sum in \C{bdyvan}. Using the fact that each $\mathcal{A}_i$ satisfies Poisson equation, we get an expression for the finite part of the amplitude. Let us consider the contribution due to $\mathcal{A}_1$ to be concrete. 
From \C{defa1}, this leads to
\bea &&\Big(\Delta_\Omega-56\Big)I_1^{finite}  \non \\ &&=-\frac{10a_4}{143} \sum' \int_0^\infty dV_2 V_2^4 \int_1^\infty \frac{d\tau_2}{\tau_2} \Big[ \frac{98}{3} + 15(\tau_2^2 +\tau_2^{-2}) \Big]e^{-\pi^2V_2\mathcal{V}_2\Big(\tau_2 \vert \hat{m}_1 +\hat{m}_2\Omega\vert^2 +\tau_2^{-1}\vert \hat{n}_1 +\hat{n}_2\Omega\vert^2\Big)/\Omega_2}.\non \\ \eea
Using the symmetry of the integral under $\tau_2 \rightarrow \tau_2^{-1}$ we get that
\bea\label{zero} &&\Big(\Delta_\Omega-56\Big)I_1^{finite}  \non \\ &&=-\frac{5a_4}{143} \sum' \int_0^\infty dV_2 V_2^4 \int_0^\infty \frac{d\tau_2}{\tau_2} \Big[ \frac{98}{3} + 15(\tau_2^2 +\tau_2^{-2}) \Big]e^{-\pi^2V_2\mathcal{V}_2\Big(\tau_2 \vert \hat{m}_1 +\hat{m}_2\Omega\vert^2 +\tau_2^{-1}\vert \hat{n}_1 +\hat{n}_2\Omega\vert^2\Big)/\Omega_2}.\non \\ \eea
Now substituting 
\be V_2\tau_2 = x, \quad V_2 \tau_2^{-1} =y,\ee
the integrals can easily be performed leading to
\be \Big(\Delta_\Omega-56\Big)I_1^{finite} = -\frac{5a_4}{4576
\pi^9} \mathcal{V}_2^{-5} \Big(294 E_{5/2}^2(\Omega,\bar\Omega) + 225 E_{3/2} (\Omega,\bar\Omega)E_{7/2}(\Omega,\bar\Omega) \Big).\ee
Exactly similar is the analysis for the expressions leading to equations for $I^{finite}_i$ for $i=2,\ldots,6$. The source term involving $\tau_2\delta(\tau_1)$ in the Poisson equation leads to $E_{5/2}^2$, while the source term involving $\tau_2 (\tau_2^2 + \tau_2^{-2})$ leads to $E_{3/2} E_{7/2}$. Note that the source terms of the second kind are actually of the form
\be \label{sym}2\tau_2 \Big( A \tau_2^2 +\frac{B}{\tau_2^2} \Big)\delta (\tau_1) = (A+B) \tau_2 \Big(\tau_2^2 +\frac{1}{\tau_2^2} \Big)\delta (\tau_1) +(A-B) \tau_2\Big(  \tau_2^2 -\frac{1}{\tau_2^2} \Big)\delta (\tau_1)\ee   
where $A$ and $B$ are constants. The first term in \C{sym} which is symmetric in $\tau_2$ leads to $E_{3/2} E_{7/2}$ in the Poisson equation. From the general structure of the analysis of the two loop supergravity amplitudes~\cite{Green:2008bf}, we expect $A=B$, and we proceed ignoring such terms\footnote{They produce source terms
\be \label{comp}\sum' \int_0^\infty dV_2 V_2^4 \int_1^\infty \frac{d\tau_2}{\tau_2} \Big(  \tau_2^2 -\frac{1}{\tau_2^2} \Big)e^{-\pi^2V_2\mathcal{V}_2\Big(\tau_2 \vert \hat{m}_1 +\hat{m}_2\Omega\vert^2 +\tau_2^{-1}\vert \hat{n}_1 +\hat{n}_2\Omega\vert^2\Big)/\Omega_2}\ee
in the Poisson equation. Its leading perturbative contribution at large $\Omega_2$ is given by setting $\hat{m}_2 = \hat{n}_2 =0$ in \C{comp} to yield
\be \mathcal{V}_2^{-5}\Omega^5 \sum_{\hat{m}_1 \neq 0, \hat{n}_1 \neq 0} \int_0^\infty dV_2 V_2^4 \int_1^\infty \frac{d\tau_2}{\tau_2} \Big(  \tau_2^2 -\frac{1}{\tau_2^2} \Big))e^{-\pi^2V_2 \Big(\tau_2 \hat{m}_1^2 +\tau_2^{-1} \hat{n}_1^2 \Big)}\ee
on simply rescaling $V_2$, which is inconsistent with string perturbation theory. }. It would be interesting to see if they vanish, and include their contributions otherwise.

Hence the Poisson equations we obtain are of the form 
\be \label{source1}\Big(\Delta_\Omega-\lambda_i\Big)I_i^{finite} \sim \frac{\mathcal{V}_2^{-5}}{
\pi^9}  \Big(E_{5/2}^2(\Omega,\bar\Omega) + E_{3/2} (\Omega,\bar\Omega)E_{7/2}(\Omega,\bar\Omega) \Big),\ee
for $i=1,2,3$ where $\lambda_i =56,30,12$ respectively, and
\be \label{source2}\Big(\Delta_\Omega-2\Big)I_4^{finite} \sim \frac{\mathcal{V}_2^{-5}}{
\pi^9}   E_{3/2} (\Omega,\bar\Omega)E_{7/2}(\Omega,\bar\Omega) ,\ee
while contributions from $\mathcal{A}_5$ and $\mathcal{A}_6$ vanish\footnote{We have that
\be I_6^{finite} \sim E_3 (\Omega,\bar\Omega) \sim\zeta(6) \Omega_2^3 +\zeta (5) \Omega^{-2} \ee
which is inconsistent with string perturbation theory and must vanish in the whole amplitude..}. This conclusion is also true for the \C{add2} term coming from the part of the amplitude coming from the ladder skeleton diagram. 

Thus they lead to the perturbative contributions given by
\bea \pi^9\mathcal{V}_2^5 I_1^{finite} &\sim &c_1 \Omega_2^8 +c_2\Omega_2^{-7}+\Big(\zeta(5)^2 +\zeta(3)\zeta(7)\Big)\Omega_2^5 +\zeta(2)\zeta(7)\Omega_2^3 \non \\ &&+ \zeta(4)\zeta(5) \Omega_2  + \zeta(3)\zeta(6)\Omega_2^{-1} + \zeta(8) \Omega_2^{-3},\non \\
\pi^9\mathcal{V}_2^5 I_2^{finite} &\sim& c_3 \Omega_2^6 +c_4\Omega_2^{-5}+\Big(\zeta(5)^2 +\zeta(3)\zeta(7)\Big)\Omega_2^5 +\zeta(2)\zeta(7)\Omega_2^3 \non \\ &&+ \zeta(4)\zeta(5) \Omega_2  + \zeta(3)\zeta(6)\Omega_2^{-1} + \zeta(8) \Omega_2^{-3},\non \\
\pi^9\mathcal{V}_2^5 I_3^{finite} &\sim &c_5 \Omega_2^4 +c_6\Omega_2^{-3}+\Big(\zeta(5)^2 +\zeta(3)\zeta(7)\Big)\Omega_2^5 +\zeta(2)\zeta(7)\Omega_2^3 \non \\ &&+ \zeta(4)\zeta(5) \Omega_2  + \zeta(3)\zeta(6)\Omega_2^{-1} + \zeta(8) \Omega_2^{-3} {\rm ln}\Omega_2,\non \\
\pi^9\mathcal{V}_2^5 I_4^{finite} &\sim& c_7 \Omega_2^2 +c_8\Omega_2^{-1}+\zeta(3)\zeta(7)\Omega_2^5 +\zeta(2)\zeta(7)\Omega_2^3 \non \\ &&+ \zeta(4)\zeta(5) \Omega_2   + \zeta(3)\zeta(6)\Omega_2^{-1}{\rm ln}\Omega_2 + \zeta(8) \Omega_2^{-3} .\eea
Of them several are ruled out by the structure of string perturbation theory, while the remaining lead to terms in the effective action given by\footnote{Some of these terms have already arisen before in \C{inconsistent}.}
\bea l_s^7 \int d^9 x \sqrt{-g^{B}} r_B^3 \Big[\zeta(4)\zeta(5) + \zeta(3)\zeta(6)e^{2\phi^B}\Big(1+{\rm ln}(e^{-\phi^B})\Big) +c_8 e^{2\phi^B} \non \\ + \zeta(8) e^{4\phi^B}\Big(1+{\rm ln}(e^{-\phi^B})\Big) + c_6 e^{4\phi^B} + c_4 e^{6\phi^B} + c_2 e^{8\phi^B}\Big]D^8 \mathcal{R}^4 +\ldots,\eea
yielding perturbative contributions upto genus five. Note that it produces logarithmically infrared divergent contributions at genus two and three.

Let us consider the contributions at genus five and four. To obtain them we need information about the coefficients $c_2$ and $c_4$ respectively. To determine them, we multiply \C{source1} by $E_8 (\Omega,\bar\Omega)$ and $E_6 (\Omega,\bar\Omega)$ respectivey and integrate over the fumdamental domain of $SL(2,\mathbb{Z})_\Omega$. This leads to~\cite{Green:2005ba}
\bea \pi^9 \mathcal{V}_2^5 c_2 &\sim &\pi^4 \sum_{k\neq 0} k^4 \int_0^\infty d\Omega_2 \Omega_2^7\Big( \mu (\vert k\vert, 5/2)^2  K_2^2 (2\pi\vert k\vert \Omega_2) \non \\ &&+ \mu (\vert k\vert, 3/2) \mu (\vert k\vert, 7/2)   K_1 (2\pi\vert k\vert \Omega_2)K_3 (2\pi\vert k\vert \Omega_2)\Big),\eea   
and
\bea \pi^9 \mathcal{V}_2^5 c_4 &\sim& \pi^4 \sum_{k\neq 0} k^4 \int_0^\infty d\Omega_2 \Omega_2^5\Big( \mu (\vert k\vert, 5/2)^2  K_2^2 (2\pi\vert k\vert \Omega_2) \non \\ &&+ \mu (\vert k\vert, 3/2) \mu (\vert k\vert, 7/2)   K_1 (2\pi\vert k\vert \Omega_2)K_3 (2\pi\vert k\vert \Omega_2)\Big), \eea
where
\be \mu(k,s) = \sum_{m>0,m\vert k} \frac{1}{m^{2s-1}}.\ee
Thus
\bea \pi^9 \mathcal{V}_2^5 c_2 \sim \pi^{-4} \sum_{k=1}^\infty \frac{1}{k^4} \Big[ \mu (k,5/2)^2 +\mu(k,3/2) \mu(k,7/2)\Big] \sim \zeta(12), \non \\ \pi^9 \mathcal{V}_2^5 c_4 \sim \pi^{-2} \sum_{k=1}^\infty \frac{1}{k^2} \Big[ \mu (k,5/2)^2 +\mu(k,3/2) \mu(k,7/2)\Big] \sim \zeta(10),\eea
on using Ramanujan's formula
\be \sum_{k=1}^\infty \frac{\mu(k,s)\mu(k,s')}{k^r} = \frac{\zeta(r)\zeta(r+2s-1)\zeta(r+2s'-1)\zeta(r+2s+2s'-2)}{\zeta(2r+2s+2s'-2)}.\ee
This leads to genus four and five contributions of the form $\zeta(8)e^{6\phi^B}r_B^3$ and $\zeta(10)e^{8\phi^B}r_B^3$ respectively, on dividing by a factor of $\pi^2$ to get the correct transcendentality. On knowing the exact coefficients, one can determine $c_6$ and $c_8$ in an analogous way\footnote{The individual contributions have a divergence as $\Omega_2\rightarrow 0$ in the integral over the fundamental domain of $SL(2,\mathbb{Z})_\Omega$.}.

\subsubsection{$O(\Lambda^6)$ counterterm contributions}

There are no contributions at this order from $I^{(c)}$. From the diagrams $e,f,g,h$ and $i$, the divergence is cancelled by a $\Lambda^6$ two loop primitive counterterm which leaves no finite remainder as discussed before. Hence there are no finite contributions that remain.

Thus to summarize, based on the structure of perturbative string theory we have obtained the contribution to the $D^8\mathcal{R}^4$ interaction at three loops coming from regularizing the ultraviolet divergences. The various loop diagrams that contribute to this amplitude are obtained from two types of skeleton diagrams: the ladder and the Mercedes skeleton diagrams. Focussing only on the divergences which can potentially yield a finite remainder, we see that the relevant one loop subdivergent contributions are $O(\Lambda^3)$ and the two loop subdivergent contributions are $O(\Lambda^6)$. On regularizing these divergences, the finite contributions come from the finite remainder of the $O(\Lambda^3)$ counterterm of one loop supergravity. It is the square of these counterterms that contribute at $O(\Lambda^6)$.

While the $O(\Lambda^6)$ divergent contribution does not leave any remainder for the diagrams that arise from the Mercedes skeleton, it does leave a remainder \C{finite} for the diagrams that arise from the ladder skeleton, which provides a part of the complete answer. This involves the $SL(2,\mathbb{Z})$ invariant coupling $E_{7/2} (\Omega, \bar\Omega)$. On the other hand, regularizing the the $O(\Lambda^3)$ divergent contributions is more involved. For the diagrams that arise from the Mercedes skeleton and some of those that arise from the ladder skeleton, we regularize by including the effects of four and five point one loop counterterms. The regularized amplitude involves performing two loop integrals, and yield Poisson equations with source terms of the form $E_{5/2}^2 (\Omega,\bar\Omega)$ and $E_{3/2} (\Omega,\bar\Omega) E_{7/2}(\Omega,\bar\Omega)$. The remaining contribution that comes from the diagram that arises from the ladder skeleton involves the square of one loop integrals leading to $E_{5/2}^2 (\Omega,\bar\Omega)$. These produce contributions upto genus five in string perturbation theory.

Thus from the analysis of the various counterterms we see that there are several possible non--vanishing contributions to the $D^8\mathcal{R}^4$ interaction from three loop quantum supergravity. Perturbatively, they lead to contributions upto genus five. It would be interesting to generalize the analysis to other non--BPS interactions at higher orders in the momentum expansion at three loops and beyond.

\section{Appendix}

\appendix

\section{The one loop subdivergence of the two loop $D^8\mathcal{R}^4$ interaction}

Ignoring an irrelevant overall numerical factor, the two loop $D^8\mathcal{R}^4$ interaction is given by
\be I^{D^8 \mathcal{R}^4} = \frac{\pi^{11} \s_4}{l_{11}^4} \sum_{\hat{m}_I \hat{n}_I} \int_0^\infty dV_2 V_2 \int_{\mathcal{F}} \frac{d^2 \tau}{\tau_2^2} e^{-\pi^2 G_{IJ} (\hat{m} +\hat{n}\tau)_I (\hat{m}+\hat{n} \bar\tau)_J V_2/\tau_2}\mathcal{B} (\tau,\bar\tau),\ee
where~\cite{Green:2008bf}
\be \mathcal{B} (\tau,\bar\tau) = \frac{4}{5} \tau_2^2 +(1-6T) +\frac{2(2-15T +40T^2)}{5\tau_2^2} +\frac{2T^2(11-43T)}{5\tau_2^4}+\frac{32T^4}{5\tau_2^6},\ee
where $T$ is given by \C{defT}. The leading ultraviolet divergence arises from a primitive two loop divergence when $\hat{m}_I =\hat{n}_I =0$ and is of the form $\Lambda^4$ coming from the boundary of the $V_2$ integral cutoff at $V_2 \sim (l_{11} \Lambda)^2$.

The subleading ultraviolet divergence comes from the boundary of moduli space when $\tau_2 \rightarrow \infty$ keeping $V_2$ fixed~\cite{Green:1999pu}. Hence we only need to isolate the boundary contribution, which is done along the lines of~\cite{Green:1999pu} and so we only mention the results. This subdivergent part is given by
\be  \Delta_\Omega I^{D^8\mathcal{R}^4}_{div}  = \frac{\pi^{11} \s_4}{l_{11}^4} \int_0^\infty d V_2 V_2 \int_{-1/2}^{1/2} d\tau_1 \Big[ \mathcal{B} \frac{\p \hat{F}_L }{\p\tau_2} - \hat{F}_L \frac{\p \mathcal{B}}{\p \tau_2}\Big]\Big\vert_{\tau_2 = (\Lambda l_{11})^2/V_2}\ee
where
\be \label{deFF} \hat{F}_L  = \sum_{\hat{m}_I, \hat{n}_I} e^{-\pi^2 G_{IJ} (\hat{m}+\hat{n}\tau)_I (\hat{m}+\hat{n}\bar\tau)_J V_2/\tau_2},\ee
and we have defined
\be \Delta_\Omega = 4\Omega_2^2 \frac{\p^2}{\p\Omega \bar\p\Omega}.\ee
This structure comes from integrating by parts and considering only the boundary contribution, as all the other contributions are finite (apart from the two loop primitive divergence). Noting that as $\tau_2 \rightarrow \infty$,
\be \mathcal{B} \rightarrow \frac{4}{5}\tau_2^2 , \quad \frac{\p \mathcal{B}}{\p \tau_2} \rightarrow \frac{8}{5} \tau_2,\ee
we get that
\be \Delta_\Omega I^{D^8\mathcal{R}^4}_{div} = -\frac{3\pi^{21/2}\s_4}{5 l_{11}}\Lambda^3 \mathcal{V}_2^{-1/2} E_{1/2} (\Omega, \bar\Omega)\ee
leading to
\be \label{Div4} I^{D^8\mathcal{R}^4}_{div} = \frac{12\pi^{21/2} \s_4}{5 l_{11}} \Lambda^3 \mathcal{V}_2^{-1/2} E_{1/2} (\Omega, \bar\Omega),\ee
leading to the term in the effective action of the form \C{1loopsubdiv}.

\section{Performing the various one loop counterterm integrals}

After factorizing out the one loop counterterms, we need to calculate several two loop integrals in the compactified theory. When the integrands in these integrals are expanded to the required order in the external momenta, we are left with integrands that involve two or four powers of the continuous loop momenta along with Lorentz scalars constructed out of the loop momenta. Thus the Lorentz structure of these integrals is fixed. Hence in these $D$ dimensional integrals, we set ($D=9$ for our case)
\be p_\mu p_\nu \rightarrow \frac{\eta_{\mu\nu} p^2}{D}, \quad p_\mu q_\nu \rightarrow \frac{\eta_{\mu\nu}p\cdot q}{D},\ee
and
\bea &&p_\mu p_\nu p_\lambda p_\rho \rightarrow \frac{1}{D(D+2)} (\eta_{\mu\nu}\eta_{\lambda\rho} + \eta_{\mu\lambda} \eta_{\nu\rho} + \eta_{\mu\rho}\eta_{\nu\lambda}) (p^2)^2, \non \\ &&p_\mu p_\nu p_\lambda q_\rho \rightarrow \frac{1}{D(D+2)} (\eta_{\mu\nu}\eta_{\lambda\rho} + \eta_{\mu\lambda} \eta_{\nu\rho} + \eta_{\mu\rho}\eta_{\nu\lambda}) p^2 (p\cdot q), \non \\ &&p_\mu p_\nu q_\lambda q_\rho \rightarrow \frac{\eta_{\mu\nu}\eta_{\lambda\rho}}{(D+2)D(D-1)}\Big((D+1)p^2 q^2 - 2 (p\cdot q)^2\Big) +\frac{\eta_{\mu\lambda} \eta_{\nu\rho} + \eta_{\mu\rho}\eta_{\nu\lambda}}{(D+2)(D-1)}\Big( (p\cdot q)^2- \frac{p^2 q^2}{D}\Big).\non \\ \eea

We now evaluate the integrals at $O(k^8)$ in \C{fige}. In $\mathcal{J}_3$
we have to keep the $O(k^2)$ term. Thus compactified on $T^2$, we get that
\bea \label{J3}
\mathcal{J}_3 &=& \frac{\pi^9S}{9(4\pi^2 l_{11}^2 \mathcal{V}_2)^2} \int_0^\infty d\s d\lambda d\rho F_L \Big[9\s\lambda \rho + 3 \s^2\lambda \rho \frac{\p}{\p\s} -\frac{\s^2\lambda^2}{4} \Big(2\frac{\p}{\p\s} - \frac{\p}{\p\rho}\Big) \Big] \Delta^{-9/2} \non \\&=& \frac{\pi^9S}{12(4\pi^2 l_{11}^2 \mathcal{V}_2)^2} \int_0^\infty d\s d\lambda d\rho F_L \s\lambda\rho \Delta^{-9/2}.\eea

In these integrals there are non--trivial factors of loop momenta in the numerator which are taken care of by replacing them with appropriate derivatives of the Schwinger parameters. For example, the integral 
\be -2\int d^{11} p d^{11}q \frac{(p\cdot k_3)(q\cdot k_4)}{p^6 q^6(p+q)^2}\ee 
in the uncompactified theory, becomes
\bea &&\frac{S}{18} \int_0^\infty d\s d\l d\r F_L \frac{\s^2\l^2}{4} \int d^9 p d^9 q \Big((p+q)^2 - p^2- q^2 \Big)e^{-\s p^2 - \l q^2 - \r (p+q)^2} \non \\ &&= -\frac{\pi^9 S}{18}  \int_0^\infty d\s d\l d\r F_L \frac{\s^2\l^2}{4}\Big(\frac{\p}{\p\r}- \frac{\p}{\p\s} - \frac{\p}{\p\l}\Big) \Delta^{-9/2}\eea
on compactifying on $T^2$.

In $\mathcal{J}_1$ and $\mathcal{J}_2$ we have to keep the $O(k^4)$ term. It is easier to write down the expression after performing the sum in \C{totcont}. 
Compactified on $T^2$, $\mathcal{J}_1$ is given by
\bea &&\sum_{S_3} S^2 \mathcal{J}_1 \non \\ &&= \frac{\pi^9\s_4}{9(4\pi^2 l_{11}^2 \mathcal{V}_2)^2}\int_0^\infty d\s d\lambda d\rho F_L \Big[ \frac{5}{12} \s^3 \lambda \frac{\p}{\p\s} +\frac{\s^4\lambda}{22} \frac{\p^2}{\p\s^2} -\frac{\s^3\lambda^2}{44} \Big(\frac{\p^2}{\p\s \p\rho} - \frac{\p^2}{\p\s \p\lambda} -\frac{\p^2}{\p\s^2}\Big)\Big]\Delta^{-9/2}\non \\ &&=-\frac{\pi^9\s_4}{12(4\pi^2 l_{11}^2 \mathcal{V}_2)^2} \int_0^\infty d\s d\lambda d\rho F_L \s^3\lambda(\lambda+\rho) \Delta^{-11/2}.\eea
Expressing the result in an $S_3$ symmetric way, we get that
\bea \label{J1}
\sum_{S_3} S^2 \mathcal{J}_1 = -\frac{\pi^9\s_4}{72(4\pi^2 l_{11}^2 \mathcal{V}_2)^2}\int_0^\infty d\s d\lambda d\rho F_L \Delta^{-9/2} \Big[ \Delta(\s+\lambda+\rho) -5\s\lambda\rho\Big].\eea

$\mathcal{J}_2$ can be calculated in exactly the same way, but the details are more involved and we omit the expression.

\section{More interactions involving one loop counterterms}

\subsection{Finite contributions from one loop counterterms}

We considered the contribution to the $D^8\mathcal{R}^4$ interaction from the counterterm in figure 4 in the main text. This involves the square of one loop amplitudes which are easily obtained leading to \C{I3}. The $D^{10} \mathcal{R}^4, D^{12} \mathcal{R}^4, D^{14} \mathcal{R}^4$ and $D^{16}\mathcal{R}^4$ interactions have two loop primitive divergences of the form $\Lambda^{11}, \Lambda^{9}, \Lambda^7$ and $\Lambda^5$ respectively, and hence receive one loop counterterm contributions in figure 4. We analyze these contributions below.   

They are obtained only from $I^{(a)}$ by expanding 
\be \label{morecal} I^{(a)} = S^4 \int \frac{d^{11}r}{r^8} \int \frac{d^{11}p}{p^2 (p+k_1)^2(p+k_1 + k_2)^2} \int \frac{d^{11}q}{q^2 (q+k_4)^2 (q+k_3 + k_4)^2}\ee
for $r\rightarrow \infty$ in the compactified theory, to the relevant order in the momentum expansion. 

Thus compactified on $T^2$, the integral 
\be \mathcal{T} = \int \frac{d^{11}p}{p^2 (p+k_1)^2(p+k_1 + k_2)^2}\ee
has to be expanded only upto $O(S^4)$. This integral gives us 
\bea \mathcal{T} = \frac{\pi^{9/2}}{4\pi^2 l_{11}^2 \mathcal{V}_2} \sum_{m_I} \int_0^\infty d\s\s^{-5/2} e^{-G^{IJ} m_I m_J \s/l_{11}^2} \int_0^1 d\omega_2 \int_0^{\omega_2} d\omega_1 e^{(1-\omega_2)(\omega_2 - \omega_1)\s S}.\eea 
The $q$ integral in \C{morecal} is the same.
On performing Poisson resummation and expanding to $O(S^4)$, this equals
\bea \mathcal{T} &=&\frac{\pi^{11/2}}{4\pi^2}\Big[ \frac{1}{2l_{11}^{5}} \Big( \frac{2}{5} (\Lambda l_{11})^5 +\frac{3}{4\pi^{9/2}} \mathcal{V}_2^{-5/2} E_{5/2} (\Omega,\bar\Omega)\Big) \non \\ &&+\frac{S}{24 l_{11}^{3}} \Big( \frac{2}{3} (\Lambda l_{11})^3 +\frac{1}{2\pi^{5/2}} \mathcal{V}_2^{-3/2} E_{3/2} (\Omega,\bar\Omega)\Big) \non \\ &&+\frac{S^2}{360 l_{11}} \Big( 2(\Lambda l_{11}) +\frac{1}{\pi^{1/2}} \mathcal{V}_2^{-1/2} E_{1/2} (\Omega,\bar\Omega)\Big) \non \\ &&+\frac{S^3 l_{11}}{6720} \Big( -\frac{2}{ (\Lambda l_{11})} +\frac{1}{2\pi^{1/2}} \mathcal{V}_2^{1/2} E_{3/2} (\Omega,\bar\Omega)\Big)\non \\ &&+\frac{S^4 l_{11}^3}{151200} \Big( -\frac{2}{ 3(\Lambda l_{11})^3} +\frac{3}{4\pi^{1/2}} \mathcal{V}_2^{3/2} E_{5/2} (\Omega,\bar\Omega)\Big)\Big] ,\eea
where we have used
\be \Gamma(s) E_s (\Omega,\bar\Omega) = \pi^{2s-1} \Gamma(1-s) E_{1-s} (\Omega,\bar\Omega)\ee
for $s=-1/2,-3/2$.
Thus the counterterm is given by
\be \delta^{(3)} I^{(a)} = \frac{c_1 \pi^3 S^4}{12 l_{11}^3 (4\pi^2)} \mathcal{T}^2,\ee
where we have to keep terms upto $O(S^8)$. Using \C{defc}, the finite contribution is given by
\bea I_4 = \frac{\pi^{16}}{18(4\pi^2)^3 l_{11}^3} \Big[ \frac{9\s_4}{32\pi^9 l_{11}^{10}} \mathcal{V}_2^{-5}  E_{5/2}^2 + \frac{\s_5}{32\pi^7 l_{11}^8} \mathcal{V}_2^{-4} E_{3/2} E_{5/2} +\frac{\s_6\mathcal{V}_2^{-3}}{240\pi^5 l_{11}^6} \Big( E_{1/2} E_{5/2} +\frac{5}{24} E_{3/2}^2\Big) \non \\ + \frac{\s_7\mathcal{V}_2^{-2}}{8960 \pi^5 l_{11}^4} \Big( E_{3/2} E_{5/2} +\frac{56\pi^2}{27}E_{1/2} E_{3/2}\Big)+\frac{\s_8 \mathcal{V}_2^{-1}}{1209600\pi^5 l_{11}^2} \Big( 9 E_{5/2}^2 + \frac{15\pi^2}{2} E_{3/2}^2 +\frac{56\pi^4}{3} E_{1/2}^2 \Big) \Big].\non \\ \eea
Dropping irrelevant numerical factors, these lead to terms in the effective action of the form (the $D^8\mathcal{R}^4$ interaction is given in \C{D8R4})
\bea l_{11}^9\int d^9 x \sqrt{-G^{(9)}} \Big[\mathcal{V}_2^{-3} E_{3/2} E_{5/2} D^{10} \mathcal{R}^4 + l_{11}^2 \mathcal{V}_2^{-2} (E_{1/2} E_{5/2} +E_{3/2}^2)D^{12} \mathcal{R}^4 \non \\ + l_{11}^4 \mathcal{V}_2^{-1} (E_{3/2} E_{5/2} +\pi^2 E_{1/2} E_{3/2})D^{14} \mathcal{R}^4 + l_{11}^6 (E_{5/2}^2 + \pi^2 E_{3/2}^2 + \pi^4 E_{1/2}^2)D^{16} \mathcal{R}^4\Big]\eea
for the various interactions. 
Dropping exponentially suppressed contributions, this leads to several perturbative contributions in the type IIB effective action of the form
\bea &&l_s^9 \int d^9 x \sqrt{-g^B} r_B \Big[ \zeta(3) \zeta(5) e^{-2\phi^B} + \zeta(2) \zeta(5) +\zeta(3) \zeta(4) e^{2\phi^B} + \zeta(6) e^{4\phi^B}\Big] D^{10} \mathcal{R}^4 \non \\ &&
+l_s^{11} \int d^9 x \sqrt{-g^B} r_B^{-1}\Big[ {\rm ln} (e^{-\phi^B}) \Big(\zeta(5) +\zeta(4)e^{4\phi^B}\Big) +\zeta(3)^2 +\zeta(2)\zeta(3) e^{2\phi^B} +\zeta(4) e^{4\phi^B}\Big] D^{12} \mathcal{R}^4 \non \\ 
&&+l_s^{13} \int d^9 x \sqrt{-g^B} r_B^{-3}\Big[ \zeta(2) {\rm ln} (e^{-\phi^B}) \Big( \zeta(3) e^{2\phi^B} + \zeta(2) e^{4\phi^B}\Big) +\zeta(3)\zeta(5) + \zeta(2)\zeta(5) e^{2\phi^B} \non \\ &&+ \zeta(3) \zeta(4) e^{4\phi^B} + \zeta(6) e^{6\phi^B}\Big] D^{14} \mathcal{R}^4\non \\
&&+l_s^{15} \int d^9 x \sqrt{-g^B} r_B^{-5}\Big[ \zeta(4) e^{4\phi^B}{\rm ln}^2 (e^{-\phi^B}) +\zeta(2) \Big(\zeta(3)^2 e^{2\phi^B} +  \zeta(2) \zeta(3) e^{4\phi^B} + \zeta(4) e^{6\phi^B}\Big) \non \\ &&+\zeta(5)^2 +\zeta(4)\zeta(5) e^{4\phi^B}+ \zeta(8) e^{8\phi^B}\Big] D^{16} \mathcal{R}^4. \eea
This leads to several perturbative contributions from genus zero to genus five for string amplitudes. The tree level and one loop $D^{10} \mathcal{R}^4$ terms agree with known results. The $D^{12} \mathcal{R}^4, D^{14} \mathcal{R}^4$ and $D^{16} \mathcal{R}^4$ interactions have infrared divergent logarithmic terms.

\subsection{Finite contributions from product of one loop counterterms}

Proceeding like the analysis above, we see that apart from the $D^8\mathcal{R}^4$ interaction, the $D^{10} \mathcal{R}^4, D^{12} \mathcal{R}^4$ and $D^{14} \mathcal{R}^4$ interactions also receive finite contributions from the counterterm in figure 5. The integrals involved are simple one loop integrals which receive contributions from $I^{(a)}, I^{(b)}$ and $I^{(d)}$ to yield (the $D^8 \mathcal{R}^4$ contribution is given in \C{totoneloop}) the counterterm
\bea \delta^{(3)}I_4 &=& \frac{9\pi^{11}}{2(4\pi^2)^3l_{11}^{11}}\Big( \frac{c_1\pi^3}{12}\Big)^2 \Big[ \frac{\s_5}{6} \Big( \frac{2}{5} (\Lambda l_{11})^5 +\frac{3}{4\pi^{9/2}} \mathcal{V}_2^{-5/2} E_{5/2} (\Omega,\bar\Omega)\Big) \non \\ &&+\frac{\s_6 l_{11}^2}{30} \Big( \frac{2}{3} (\Lambda l_{11})^3 +\frac{1}{2\pi^{5/2}} \mathcal{V}_2^{-3/2} E_{3/2} (\Omega,\bar\Omega)\Big) \non \\ &&+\frac{\s_7 l_{11}^4}{840} \Big( 2(\Lambda l_{11}) +\frac{1}{\pi^{1/2}} \mathcal{V}_2^{-1/2} E_{1/2} (\Omega,\bar\Omega)\Big)\Big],\eea 
leading to finite contributions given by
\bea \label{needref}I_4 = \frac{\pi^{15}}{4608 l_{11}^{11}} \Big[ \frac{\s_5}{8\pi^{9/2}}\mathcal{V}_2^{-5/2} E_{5/2} + \frac{\s_6 l_{11}^2}{60\pi^{5/2}} \mathcal{V}_2^{-3/2} E_{3/2} + \frac{\s_7 l_{11}^4}{840\pi^{1/2}}\mathcal{V}_2^{-1/2} E_{1/2}\Big] \eea
on using \C{defc}.
Now \C{needref} leads to terms in the effective action of the form
\be l_{11}^9 \int d^9 x \sqrt{-G^{(9)}} \Big[ \mathcal{V}_2^{-3/2} E_{5/2} D^{10} \mathcal{R}^4 + l_{11}^2\mathcal{V}_2^{-1/2} E_{3/2} D^{12} \mathcal{R}^4 + l_{11}^4\mathcal{V}_2^{1/2} E_{1/2} D^{14} \mathcal{R}^4\Big] \ee
which produces terms of the form
\bea &&l_s^9 \int d^9 x\sqrt{-g^B} r_B^{-1} \Big[\zeta(5) +\zeta(4)e^{4\phi^B}\Big] D^{10}\mathcal{R}^4 \non \\ &&+l_s^{11} \int d^9 x\sqrt{-g^B} r_B^{-3} \Big[\zeta(3) e^{2\phi^B}+\zeta(2) e^{4\phi^B}\Big] D^{12} \mathcal{R}^4 \non \\ &&+l_s^{13}\int d^9 x\sqrt{-g^B}r_B^{-5} e^{4\phi^B}{\rm ln}(e^{-\phi^B})D^{14} \mathcal{R}^4\eea
in the type IIB effective action on dropping exponentially suppressed corrections.




\providecommand{\href}[2]{#2}\begingroup\raggedright\endgroup

\end{document}